\documentclass[aps,pre,twocolumn,showpacs,groupedaddress]{revtex4}
\usepackage{graphicx,color}
\begin{document}

\title{Structure of finite sphere packings via exact enumeration: Implications for colloidal crystal nucleation}
\author{Robert S. Hoy$^{1,2}$}
\author{Jared Harwayne-Gidansky$^{3,4}$}
\author{Corey S. O'Hern$^{1,2,4}$}
\affiliation{Departments of Mechanical Engineering \& Materials Science$^1$, Physics$^2$, Electrical Engineering$^3$, and $^4$Integrated Graduate Program in Physical and Engineering Biology, Yale University, New Haven, CT}
\date{\today}
\begin{abstract}
We analyze the geometric structure and mechanical stability of a
complete set of isostatic and hyperstatic sphere packings obtained
via exact enumeration.  The number of nonisomorphic isostatic packings
grows exponentially with the number of spheres $N$, and their
diversity of structure and symmetry increases with increasing $N$ and
decreases with increasing hyperstaticity $H \equiv N_c - N_{ISO}$,
where $N_c$ is the number of pair contacts and $N_{ISO} = 3N-6$.
Maximally contacting packings are in general neither the densest nor
the most symmetric.  Analyses of local structure show that the
fraction $f$ of nuclei with order compatible with the bulk (RHCP)
crystal decreases sharply with increasing $N$ due to a high propensity
for stacking faults, 5- and near-5-fold symmetric structures, and
other motifs that preclude RHCP order.  While $f$ increases with
increasing $H$, a significant fraction of hyperstatic nuclei for $N$
as small as $11$ retain non-RHCP structure.  Classical theories of
nucleation that consider only spherical nuclei, or only nuclei with
the same ordering as the bulk crystal, cannot capture such effects.
Our results provide an explanation for the failure of classical
nucleation theory for hard-sphere systems of $N\lesssim 10$ particles;
we argue that in this size regime, it is essential to consider nuclei
of unconstrained geometry.  Our results are also applicable to
understanding kinetic arrest and jamming in systems that interact via
hard-core-like repulsive and short-ranged attractive interactions.
\end{abstract}
\maketitle

\section{Introduction}

Crystallization of monodisperse hard spheres is a complex problem for
several reasons.  In the absence of attractive interactions,
crystallization is associated with minimization of free volume.  For
bulk systems, the ground states, the FCC and HCP lattices, as well as
other stacking variants of hexagonal planes, possess volume fraction
$\phi_{\rm xtal} = \pi/\sqrt{18} \simeq .7405$~\cite{hales98}.  In
addition, there are an exponential number of rigid packings with
volume fractions that range from random close packing~\cite{berryman83} 
to $\phi_{\rm xtal}$.  The large number of metastable structures and
large barriers separating the amorphous and crystalline states leads
to formation of amorphous structures
\cite{rintoul96,torquato00,ohern03,torquato10,foffi00} if the quench
rate is not sufficiently slow.  This makes understanding crystal
\textit{nucleation} in these systems particularly important since
jamming or glass-formation may be avoided only through nucleation and
growth of crystallites.  Since the bulk crystal state maximizes
$\phi$, classical nucleation theory suggests that the densest packings
of $N$ spheres within a (minimal) spherical volume $V$ are optimal
nuclei.  However, this approach should work if and only if these
packings possess the same structural order as the bulk crystalline
phase.  A recent study by Hopkins \textit{et al.}\ \cite{hopkins11}
showed that this condition fails; the densest packings did not in
general have FCC, HCP, or Barlow \cite{barlow1883,barlowfoot} order.
Instead, their surface order was dominated by the spherical boundary
conditions, and so they may not correspond to the stable nuclei that
form in unconfined geometries, \textit{i.e.}\ within an arbitrary
volume in a larger system.

It is important to note that small nuclei may be distinctly
\textit{aspherical}, with a wide variety of shapes, symmetries, and
formation probabilites.  Specifically, $N$-sphere nuclei with $N_c$
contacts can form $\mathcal{M}(N,N_c)$ packings of distinguishable
shape, symmetry, and entropy \cite{meng10}.  However, the full range
of shapes, symmetries and statistical-gometrical properties of such
packings has not been quantitatively characterized, even for small
$N$.  There have been numerous studies of the phase diagrams and
crystal nucleation and growth in systems of hard spheres and sticky
hard spheres
~\cite{miller04,schilling10,taffs10,omalley03,zaccarrelli09,karayiannis11}.
However, studies of hard sphere crystal nucleation in particular have
shown that quantities inferred from experimental results and classical
nucleation theory can differ from simulation results \cite{auer01} by
orders of magnitude.  Thus, there is a need to characterize the
structural properties of nuclei posessing arbitrary geometry and interactions with the 
surrounding fluid \cite{miller04,crocker10} to gain a more quantitatively accurate description of crystal
nucleation and growth.  In this paper, we analyze the
statistical-geometrical properties of small sticky-hard-sphere nuclei
to lay a foundation for a more quantitative understanding of crystal
nucleation in systems such as colloids with hard-core-like repulsions
and short-range attractions \cite{buzzacaro07,meng10}.

Determining these properties is largely an exercise in the geometry of
finite sphere packings
\cite{hoare76,arkus09,bernal60,conway98,arkus11}.  The two key
mathematical problems are ``What is the maximum number of contacts
$N_c^{max}(N)$ that can be formed by $N$ monodisperse spheres?'' and
``How many different ways $\mathcal{M}(N, N_c)$ can $N$ spheres form
$N_c$ contacts?''  Solving these problems simultaneously yields
complete sets of isocontacting packings of hard spheres and of
isoenergetic states of sticky hard spheres.  These in turn have
applications to physical problems ranging from crystallization and
jamming
\cite{rintoul96,torquato00,foffi00,auer01,ohern03,omalley03,zaccarrelli09,torquato10,karayiannis11,schilling10}
to cluster physics \cite{wales04,arkus09,meng10} to liquid structure
\cite{stell91,mossa03,taffs10,hhgw11} to protein folding
\cite{richards77,tenwolde97}, as well as engineering applications such
as circuit design \cite{miller97} and error-correcting codes
\cite{johnson62}.

Despite this wide applicability, progress in obtaining solutions has
been slow.  Determining $N_c^{max}(N)$ corresponds to the generalized
Erd\H{o}s unit distance problem \cite{erdos46} in three dimensions,
which remains unsolved, while determining $\mathcal{M}(N, N_c)$ has
been proven to be algorithmically ``NP-complete'' \cite{horvat11}.
The latter condition impedes calculation of $\mathcal{M}(N, N_c)$ via
Monte Carlo or related methods \cite{footMC}.  Consequentially,
$\mathcal{M}(N, N_c^{max}(N))$ has been determined for $N$ as large as
$10$ only recently \cite{arkus09,hoy10,arkus11}.  Here we present an
efficient method for finding $N_c^{max}(N)$, $\mathcal{M}(N,N_c)$ and
the permutational entropies of distinguishable sphere packings.
$\mathcal{M}$ is an integer for isostatic ($N_c = 3N-6 \equiv
N_{ISO}$) and hyperstatic ($N_c > 3N-6$) clusters of sticky spheres
\cite{hoare76,arkus09} wherein each sphere possesses at least 3
contacts.  We focus on packings satisfying these necessary
\cite{footfloppy} conditions for mechanical stability since they
correspond to solidlike clusters that likely play an important role in
nucleation and growth of crystals.

We find all isostatic and hyperstatic sticky hard sphere packings for
$N \leq 11$ via exact enumeration, and present novel analyses of
several statistical-geometrical properties of this complete set of
packings that are relevant for understanding crystallization and
jamming.  Several dramatic features are associated with the increasing
maximum hyperstaticity $H_{max}(N) = N_c^{max}(N) - N_{ISO}$ for $N >
9$.  Key amongst these are that maximally contacting packings are in
general quite different from the densest packings.  Minimal energy
(maximally contacting) packings are not necessarily either the most
compact or symmetric.  Instead, the most symmetric and compact
packings are often mechanically stable ``excited states'' with $N_c <
N_c^{max}(N)$.  While many of these stable packings correspond to
``on-pathway'' nuclei possessing structural order consistent with the
bulk (Barlow-ordered) crystalline phase, many do not.  The latter
correspond to `off-pathway'' nuclei possessing structural motifs
incompatible with Barlow order, such as 5-fold-symmetries, stacking
faults, and twin defects, all of which are known to impede
crystallization~
\cite{buzzacaro07,omalley03,zaccarrelli09,karayiannis11,schilling10}.
The fraction of isostatic nuclei possessing non-Barlow order grows
rapidly with increasing $N$ to nearly 95\% for $N=11$.  Crystalline
order increases with increasing hyperstaticity, yet a significant
fraction ($\sim 50\%$) of hyperstatic nuclei for $N$ as small as $11$
retain non-Barlow structure.

The outline of the remainder of the manuscript is as follows.  In
Section \ref{sec:methods} we describe our exact enumeration procedure,
focusing particularly on advances beyond those employed in previous
studies \cite{arkus09,hoy10,arkus11}.  Section \ref{sec:results}
presents analyses of the structure and symmetry of the sphere
packings, including size, shape, and the statistical prominence of key
structural motifs.  Section \ref{sec:discussion} relates our work to
other studies of crystal nucleation that include interactions of
nuclei with the bulk fluid and a range of nucleation pathways.  In
Section \ref{sec:conclude}, we place our results in context with other
recent work, and conclude.  Finally, Appendix \ref{sec:bridge}
describes mechanical stability analyses of the nuclei,
Appendix~\ref{sec:solvervalidation} shows convergence of the exact
enumeration procedure, and Appendix~\ref{sec:implic} explains implicit
contact graphs for sphere packings.

\section{Methods}
\label{sec:methods}

\begin{figure}[htbp]
\includegraphics[width=2.5in]{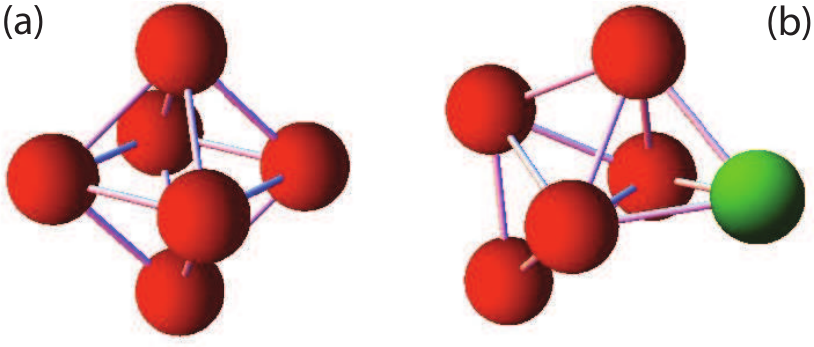}
\caption{(Color online) Macrostates: $\mathcal{M}(6,12) = 2$.  The
octahedral structure (a) has high symmetry and low permutational
entropy ($\omega_a = 15$), while the capped trigonal bipyramid
structure (b) has low symmetry and high permutational entropy
($\omega_b = 180$) \cite{meng10}.  The position of the green
(rightmost) sphere in (b) implies a stacking fault.  Note that in this
and many subsequent figures, sphere sizes are reduced for visual
clarity, and the connecting  bars indicate pair contacts.}
\label{fig:macrostateexamp}
\end{figure}

A key feature of (sticky) hard sphere packings is that their
(isoenergetic) isocontacting states are in general highly degenerate.
Many distinguishable arrangements (``macrostates'') of $N$ particles
with $N_c$ contacts are possible.  Here we employ a particular
definition of the term \textit{macrostate}.  An $N$-particle,
$N_c$-contact macrostate is defined by a unique set of $N(N-1)/2$
squared interparticle distances $\{r_{ij}^2\}$
\cite{footfloppy,footchiral}.  Different macrostates have different
$\{r_{ij}^2\}$ and ``shapes''.  In general, systems of $N$ spheres
with $N_c$ contacts possess $\mathcal{M}(N,N_c)$ distinguishable
macrostates.  For example, $\mathcal{M}(6,12) = 2$ since exactly two
six-particle macrostates exist for systems with $N_c = N_c^{max}(6) =
12$ (Fig.\ \ref{fig:macrostateexamp}) \cite{hoare76}.

There are many ways to organize indistinguishable spheres into any
given macrostate; these correspond to permutations of particle indices
$\{i,j\}$ that preserve $\{r_{ij}^2\}$.  We refer to the number of
allowed permutations as the number of \textit{microstates} $\omega_k$
corresponding to a particular macrostate $k$.  Differing structure and
symmetry of macrostates imply they have different $\omega_k$
(\textit{i.e.}\ permutational entropies).  For example, the highly
symmetric octahedral structure shown in Fig.\
\ref{fig:macrostateexamp}(a) has $\omega_a = 15$, while the less
symmetric structure shown in panel (b) has $\omega _2 = 180$.  Note
that (a) is a subset of the FCC and HCP lattices, while the capped
trigonal bipyramid (b) is a stack-faulted structure; for example, if
the green sphere is removed, the remaining 5 spheres have HCP order.
Such effects have important implications for nucleation, as
macrostates with higher $\omega$ will form with greater probability
\cite{meng10,wales10}.  These will be discussed in detail below.

The potential for sticky hard spheres with diameter $D$ and contact attractions is \cite{yuste93}:
\begin{equation}
U_{ss}(r) = \Bigg{\{}\begin{array}{ccc}
\infty & , & r < D\\
-\epsilon & , & r = D\\
0 & , & r > D.\\
\end{array}
\label{eq:stickyspherepot}
\end{equation}
Since $U_{ss}(r)$ possesses no scale, and we consider monodisperse
systems, the value of $D$ is arbitrarily set to unity below.
Hard-sphere constraints imply that the center-to-center distances
$r_{ij}$ between unit spheres $i$ and $j$ with positions $\vec{r}_i$
and $\vec{r}_j$ obey $r_{ij} \geq 1$, where the equality holds for
contacting pairs.

In this manuscript, we enumerate the global and low-lying local
potential energy minima~\cite{footfloppy} of Eq.\
\ref{eq:stickyspherepot}.  It is worth noting that at finite
temperature $T$ and zero pressure, systems interacting via the sticky
hard-sphere potential (Eq.\ \ref{eq:stickyspherepot}) will have no
persistent contacts since the range of attractive interactions is
exactly zero.  For this reason Baxter \cite{baxter68} introduced the
adhesive hard sphere (AHS) interaction potential
\begin{equation}
U_{AHS}(r) = \bigg{\{}\begin{array}{ccc}
\infty & , & r < D\\
\log\left(12\tau\displaystyle\frac{r-r_c}{D}\right) & , & D < r < r_c\\
0 & , & r > r_c\\
\end{array}
\label{eq:baxterpot}
\end{equation}
where $\tau$ is a temperature-like parameter and $r_c$ is the
interaction range.  The packings reported in this work have
configurations ($\{\vec{r}\}$) that are identical to the corresponding
AHS packings in the $\tau\to 0$ and $r_c \to 0$ limits.  While AHS systems
have also been shown to possess thermodynamic anomalies in the $r_c
\to 0$ limit \cite{stell91}, several theoretical studies
\cite{foffi00,yuste93,miller04} have shown that these vanish when
$r_c$ is as small as $\sim .01D$, \textit{e.g}\ by identifying Eq.\
\ref{eq:baxterpot} as the short-range limit of the attractive
square-well potential.  Applicability of our studies to systems
interacting via short (but finite) range potentials is discussed in
Section \ref{subsec:applic}.

\begin{figure}[h]
\includegraphics[width=3.35in]{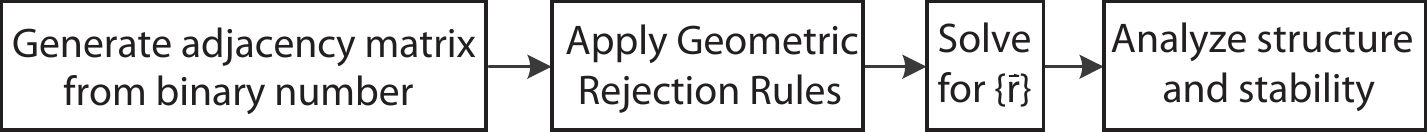}
\caption{Schematic of our exact enumeration algorithm including structural and stability analyses.}
\label{fig:topschem}
\end{figure}

Employing an infinitely narrow potential well also allows all
isostatic and hyperstatic $N$-sphere configurations to be conveniently
characterized by $N\times N$ \textit{adjacency matrices} $\bar{A}$
with $A_{ij} = 1$ for contacting particles and $A_{ij} = 0$ otherwise.
In general, the configuration $\{\vec{r}\} = \{\vec{r}_1, \vec{r}_2,
... \vec{r}_N\}$ can be solved (\textit{cf.}\ Section
\ref{subsec:euclidean}) from $\bar{A}$ \cite{footfloppy} provided two
minimal conditions for mechanical stability are met \cite{jacobs95}:
\textbf{(i)} each particle possesses at least 3 contacts and
\textbf{(ii)} $N_c \geq N_{ISO}$.  Throughout this paper, we refer to
these as conditions \textbf{(i)} and \textbf{(ii)}.  Using an
efficient exact enumeration algorithm schematically depicted in Fig.\
\ref{fig:topschem}, we identify all $\bar{A}$ and $\{\vec{r}\}$ that
correspond to nonoverlapping configurations with $N_c \geq N_{ISO}$
contacts.  These configurations all have equal potential energy $U =
-\epsilon \sum_{j > i} A_{ij} = -N_c\epsilon$ for sticky spheres, and
their $\{\vec{r}\}$ are \textit{identical} to corresponding
$N_c$-contact hard sphere packings (to see this, consider approaching
the limit $\epsilon \to 0$ from $\epsilon < 0$).  Overlapping
configurations with one or more $r_{ij} < 1$ have infinite $U$.  We
refer to configurations with no interparticle overlaps as ``valid
packings'' and overlapping configurations as ``invalid packings.''

The rest of this section describes our exact enumeration method in detail, following the scheme depicted in Fig.\ \ref{fig:topschem}.  
For each $N$ and $N_c$, we perform complete enumeration by efficiently iterating over all adjacency matrices (Section \ref{subsec:exactpolymer}).
Valid packings are found by applying geometrical and graph-theoretic rejection rules (Sections \ref{subsec:geomrejec} and \ref{subsec:systematic}) and then solving (Section \ref{subsec:euclidean}) for the structure $\{\vec{r}\}$ of nonisomorphic packings passing these rejection rules.
The structure and stability of valid packings are then analyzed as described in Sections \ref{subsec:dynmat} and \ref{subsec:structanaly}.

\subsection{Exact Enumeration Method}
\label{subsec:exactpolymer}

Since all elements of adjacency matrices are 0 or 1, they correspond to binary numbers $\mathcal{B}$.
The matrices are symmetric, and diagonal elements are zero by convention, so any $\bar{A}$ may be uniquely associated with one of $2^{N(N-1)/2}$ distinct $\mathcal{B}$.
All sticky sphere packings may be found by iterating sequentially  \cite{footbinary} over the $\mathcal{B}$ and mapping each to an adjacency matrix .

The number of adjacency matrices that must be iterated over to find all macrostates and microstates for fixed $N$ and $N_c$ but with no constraints on the arrangement of the elements (\textit{i.e.}\ arbitrary topology) is
\begin{equation}
\mathcal{N}_{arb} = \displaystyle\frac{[(N^{2} - N)/2]!}{N_c![(N^{2} - N)/2 - N_c]!}.
\label{eq:constrainedcontact}
\end{equation}
$\mathcal{N}_{arb}$ grows faster than $\exp(N)$ and rapidly becomes prohibitively large; for example, $\mathcal{N}_{arb} = 3.824\cdot 10^{15}$ for $N = 11$ and $N_c = 27$. 

\begin{figure}[h]
\includegraphics[width=3in]{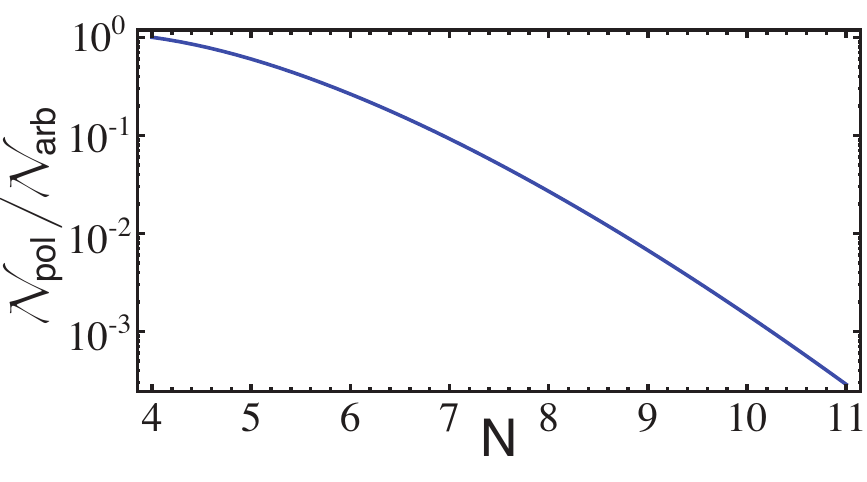}
\caption{(Color online) Computational effort reduction obtained by using polymeric topology in enumeration.  The graph shows faster than exponential decay of $\mathcal{N}_{pol}/\mathcal{N}_{arb}$ with increasing $N$ for $N_c = 3N-6$.}
\label{fig:polymerspeedup}
\end{figure}

\begin{figure}[htbp]
\centering
\includegraphics[width=2.7in]{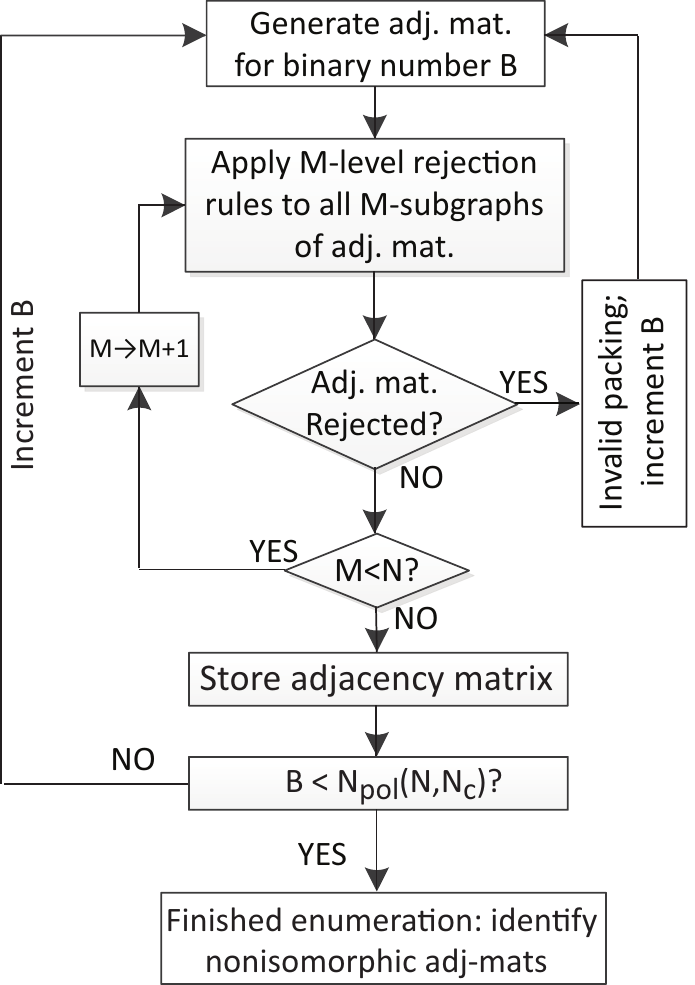}
\caption{Schematic of Part 1 of our enumeration protocol: Geometric rejection rules. The sequential enumeration over binary numbers and application of rejection rules correspond to the left two boxes in Fig.\ \ref{fig:topschem} and are implemented as described in Sections \ref{subsec:exactpolymer}-\ref{subsec:systematic}.}
\label{fig:algschematic}
\end{figure}

However, all macrostates can be found with greatly reduced computational effort through appropriate selection of ``topological'' constraints on the elements of $\bar{A}$.
Biedl \textit{et al.} \cite{biedl01} proved that all connected sphere packings admit linear polymeric paths, \textit{i.e.}\ for any valid packing, one can always permute particle indices so that the packing is fully traversed by a ``polymeric'' $\bar{A}$ with $A_{i,i+1} = 1$ for all $i$.
As in Ref.\ \cite{hoy10}, we impose polymer topology by fixing $A_{i,i+1} = 1$.
Thus $N-1$ elements of $\bar{A}$ are fixed to unity, and the remaining $N(N-1)/2 - (N-1)$ elements are left unconstrained.
This arrangement reduces the number of binary numbers and adjacency matrices over which one must iterate to
\begin{equation}
\label{constraint}
\mathcal{N}_{pol} = \displaystyle\frac{[(N^{2} - 3N + 2)/2]!\ ([N_c - (N-1)]!)^{-1}}{[(N^{2} - 3N + 2)/2 - ((N_c - (N-1))]!}.
\label{eq:constrainedpermutations}
\end{equation}
For each $N$ and $N_c$, we iterate sequentially over the $\mathcal{N}_{pol}$ binary numbers and adjacency matrices as illustrated in Fig.\ \ref{fig:algschematic}.
$\mathcal{N}_{pol}/\mathcal{N}_{arb}$ decreases faster than exponentially with increasing $N$, with a corresponding reduction of computational effort.
For the purposes of calculating $\mathcal{M}(N, N_c)$, this provides a speedup of about 3 orders of magnitude for the largest $N$ considered here (Fig.\ \ref{fig:polymerspeedup}.)

Here we are considering colloidal clusters with no fixed topology.
Enumerating over adjacency matrices with polymer topology naturally produces $\{\omega_k\}$ corresponding to packings with polymer topology, \textit{i.e.}\ different absolute macrostate populations $\omega_k$ and ratios $\omega_k/\omega_j$ ($j,k \in \{1,\mathcal{M}\}$) due to entropic factors such as blocking \cite{hoy10}.
However, the $\omega_k$ for colloidal clusters may be calculated via symmetry operations:
\begin{equation}
\omega_k = C_k N!/\mathcal{A}_k
\label{eq:omegak}
\end{equation}
where $\mathcal{A}_k$ is the number of automorphisms of the adjacency matrix corresponding to macrostate $k$ \cite{footAk}. 
$C_{k} = 2$ for macrostates possessing chiral enantiomers and $1$ for those which do not \cite{arkus09,footchiral}. 
The total number of microstates for nuclei with N particles and $N_c$ contacts is then 
\begin{equation}
\Omega(N, N_c) = \sum_{k = 1}^{\mathcal{M}(N, N_c)} \omega_k.
\label{eq:omega}
\end{equation}
Note that while Eq.\ \ref{eq:omegak} treats particles $\{1,N\}$ as distinguishable \cite{footdisting}, we have verified that Eqs.\ \ref{eq:omegak}-\ref{eq:omega} produce the same $\omega_k$ and $\Omega$ produced by an alternative method \cite{meng10,arkus11} that treats particles as indistinguishable and calculates $\omega_k$ using the number of symmetries $c$ possessed by the coordinate solutions $\{\vec{r}_k\}$.

\subsection{Geometric Rejection Rules}
\label{subsec:geomrejec}

Valid sticky sphere packings correspond to $N$-vertex, $N_c$-contact unit distance graphs that are embeddable  \cite{horvat11} in three dimensions.
A key advance in determining embeddability of small packings was recently made by Arkus, Brenner and Manoharan \cite{arkus09,arkus11}.
They used concepts from sphere geometry to develop \textit{geometric rejection rules} identifying invalid packings with patterns within adjacency matrices.
Geometric rejection rules facilitate connections to graph theory and enable formulation of rules in terms of Boolean satisfiability conditions.
These conditions can be conveniently organized into a series of ``M''-rules that reject an invalid $\bar{A}$ based on patterns within $M\times M$ subgraphs of  $\bar{A}$, or restrict the way additional spheres can be added to form valid ($M+1$)-particle packings.
We apply these in order of increasing $M$ to reject invalid packings.

As illustrated in Fig.\ \ref{fig:algschematic}, we apply the rejection rules to all 
\begin{equation}
F(M) =  \displaystyle\frac{N!}{M!(N-M)!},
\label{eq:ruleordereffort}
\end{equation}
$M\times M$ subgraphs of each $\bar{A}$.  
Any subgraph violating any rule indicates an invalid packing.
For example, at each rule-level $M$, we test for over-connected ``O''-clusters as follows. 
$O$-clusters are defined by the $M$-subgraph and the $O-M$ particles that contact at least one particle in the $M$-subgraph.  
We independently verfied that $N_c^{max}(O)$ is the same as found in Refs.\ \cite{hoare76,arkus09} for $O \leq 11$.
If the total number of contacts in the $O$-cluster is greater than $N_c^{max}(O)$ (\textit{i.e.}\ $3O-6$ for $O \leq 9$, $3O-5$ for $O = 10$, and $3O-4$ for $O=11$), we reject the adjacency matrix.    These overconnected-subcluster rules eliminate many invalid packings and improve the efficiency of the code. 

Refs.\ \cite{arkus09,arkus11} reported a complete set of rejection rules for packings of $N \leq 7$ spheres.
This is equivalently a complete set of rejection rules for $M \leq 7$ subclusters (with $M_c = 3M-6$ contacts) within larger packings.
We have extended this set to reject all invalid isostatic and hyperstatic packings of $N \leq 9$ particles, and many invalid packings of $N > 9$ particles.
Note that in contrast to Ref.\ \cite{arkus11}, we do \textit{not} use the triangular bipyramid rule \cite{footribi}, nor explicitly search for conflicts in the distance matrix $\bar{D}$ ($D_{ij} = r_{ij}$) arising from different $M < N$ subgraphs.  
Instead we employ geometric and graph-theoretic rejection rules that do not require calculation of unknown distances.
Space constraints preclude describing our complete set of rejection rules for $M \leq 9$; here we highlight several rules not contained in Refs.\ \cite{arkus09,arkus11}.

Several rejection rules are obtained from known graph-theoretic results for the embeddability of sphere packings.
Kuratowski graphs \cite{kuratowski} $K_{m,n}$ have $m+n$ vertices and $mn$ edges, with each of the vertices having degree $n$.
The graphs $K_{3,3}$, $K_{3,4}$, and $K_{5,4}$ are not embeddable as 3D sphere packings \cite{hlinenythesis,harborth02}; these results imply rejection rules for ($M = 6,\ M_c = 9$), ($M = 7,\ M_c = 14$), and ($M = 9,\ M_c = 18$) subgraphs, respectively.

\begin{figure*}
\includegraphics[width=6.5in]{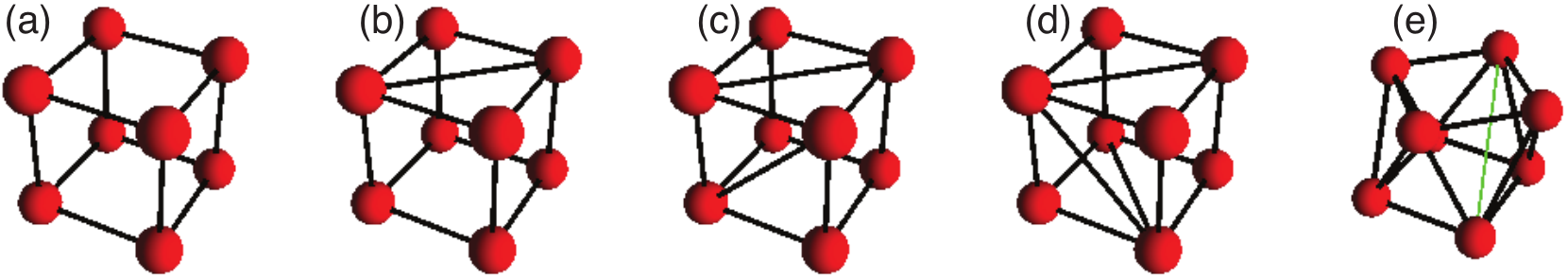}
\caption{(Color online) Schematic for $M=8$ cube and sheared cube-based geometric rejection rules.  No 9th particle may contact more than four particles of any subgraph isomorphic to the cube (panel (a)), sheared cubes with contacts across one (panel (b)), two (panel (c)), or three (panel (d)) faces.  Similar exclusions apply to other subgraphs similar to panels (c-d) but with different topology, \textit{e.g.}\ two cross-face contacts on opposite as opposed to adjacent faces.  Panel (e): no 9th particle may contact more than four particles of the square antiprism with $M_c = 16$ or the sheared square antiprism with $M_c=17$ (the 17th contact is indicated by the green line).}
\label{fig:shearedcubes}
\end{figure*}

We make use of the fact that no structures with BCC symmetry are among the isostatic or hyperstatic packings for sticky hard spheres \cite{footoctexcep} since placing a ninth sphere inside a cube implies overlap.
Figure \ref{fig:shearedcubes}(a) shows a cubic structure with $M=8$ and $M_c = 12$.
Placing a ninth sphere in the interior of the cube to form a putative $N=9$, $N_c = 20$ BCC packing implies an overlap of at least $(\sqrt{3}-1)/4$, \textit{i.e.}\ an $r_{ij} \leq 1 - (\sqrt{3}-1)/4$).
As illustrated in Fig.\ \ref{fig:shearedcubes}, many $M=8$ rejection rules are obtained by observing that a ninth sphere must lie in the interior of a cube or sheared cube if it contacts more than 4 of the 8 particles; any such placement implies particle overlap.  These rules eliminate many invalid $N=9, N_c = 21$ packings not eliminated by $M < 8$ rules, and become increasingly effective at eliminating invalid packings for $N > 9$.

\begin{figure}
\includegraphics[width=2.5in]{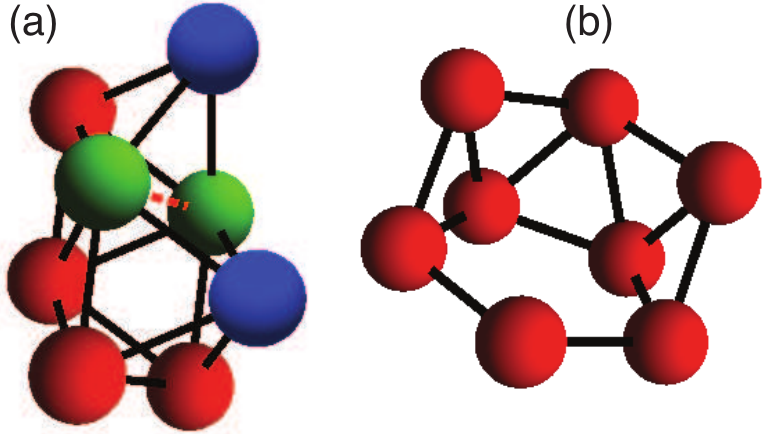}
\caption{(Color online) ``Irregular'' $M=8$ rules.  Panel (a): Contact of a ninth monomer with the red and blue (darkest shaded) monomers implies overlap of the green (lightest shaded) monomers (red dashed line, $d_{gg} < 1$) and/or a planar angle $\Psi < 2\pi/3$ (see Fig.\ \ref{fig:open4ringang}). Panel (b): Contact of a ninth monomer with each of the eight shown implies at least one overlap.}
\label{fig:M8irregular}
\end{figure}

Other $M = 8$ rejection rules relate to ``irregular'' seeds lacking any underlying cube or sheared-cube topology.
Figure \ref{fig:M8irregular}(a) shows an $M = 8$, $M_c = 17$ packing that cannot be ``4-kissed'' to form a 9/21 packing.  A ninth monomer cannot contact the four (blue and green) monomers because doing so would imply overlap between the green monomers (\textit{i.e.}\ an interparticle distance $d_{gg} \simeq 0.615$), indicated by the dashed red line.  
Figure \ref{fig:M8irregular}(b) shows an $M = 8$, $M_c = 13$ packing with the topology of a partial icosohedron.  A ninth monomer cannot contact all eight to form a $N = 9, N_c = 21$ packing without implying overlap.
The rules shown in Figs.\ \ref{fig:shearedcubes}-\ref{fig:M8irregular}, together with a few additional similar rules for $M < 8$, are sufficient to reject all invalid $N = 9,\ N_c = 21$ packings.

\subsection{Systematic development of additional rejection rules}
\label{subsec:systematic}

\begin{figure}
\includegraphics[width=2.25in]{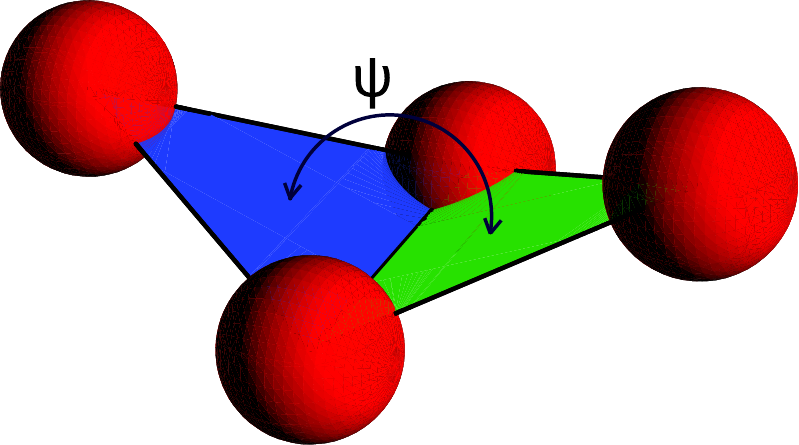}
\caption{(Color online) A useful rejection rule for $M$-particle packings containing open 4-rings on their surfaces is that an (M+1)$^{st}$ particle can contact each of the 4 particles in the ring only if the planar angle $\psi$ (indicated by the arrows) satisfies $2\pi/3 \leq \psi \leq \pi$.}
\label{fig:open4ringang}
\end{figure}

The number $\mathcal{Q}$ of rejection rules required to reject all invalid $N$-sphere packings is expected to grow exponentially with $N$ \cite{footribi,footWein}.
Development of geometric rejection rules ``by hand'', as described in the above subsection, therefore becomes increasingly difficult as $N$ increases.
Here we report a systematic method for developing additional rejection rules.
We employ a ``deep-seed'' elimination procedure: 

1) Find all nonisomorphic ``$M$-seed'' graphs of $M$ vertices and $M_c < 3M-6$ edges satisfying minimal rigidity condition \textbf{(ii)} and passing all $L \leq M$ rejection rules.

2) Determine which of these can form known-valid packings of $P=M+1$ particles and $P_c$ contacts by examining all possible arrangements wherein the $P^{th}$ sphere contacts ($P_c - M_c$) spheres of the $M$-seed.

3) A seed graph that can never have such an ($P_c - M_c$) ÒkisserÓ often yields a novel rejection rule.  

For example, we find 540 nonisomorphic $M = 9$, $M_c = 20$ seed graphs that satisfy condition \textbf{(ii)} and pass all $M \leq 8$ rejection rules.
A tenth sphere can contact four (of the nine) particles to form a $P=10$, $P_c = 24$ packing for only 197 of these seeds.
The remaining 343 seeds cannot be a subgraph of any valid 10/24 packing.
Therefore all $\bar{A}$ containing subgraphs $\bar{A}'$ isomorphic to any of these 343 and a 10th particle contacting four of the particles in $\bar{A}'$ correspond to invalid packings and are rejected at the $M=9$ level.
This ``no-4-kisser'' rule is particularly effective, eliminating 132 nonisomorphic invalid $N = 10$, $N_c = 24$ packings that passed all previously implemented rules.
Most of these eliminated packings are invalid because the putative 10th sphere contacts an open 4-ring on the surface of a 9-sphere seed.
The additional four contacts can be formed without producing overlap only if the planar angle $\psi$ satisfies $2\pi/3 \leq \psi \leq \pi$ (Fig.\ \ref{fig:open4ringang}).
While all $M = 9$, $M_c = 20$ packings fail to satisfy condition \textbf{(i)}, many are insufficiently floppy for $\psi$ to fall within this range, \textit{i.e.}\ addition of a 10th sphere implies $\psi < 2\pi/3$ or $\psi > \pi$.

\subsection{Efficient Euclidean structure solver}
\label{subsec:euclidean}

\begin{figure}[h]
\centering
\includegraphics[width=2.7in]{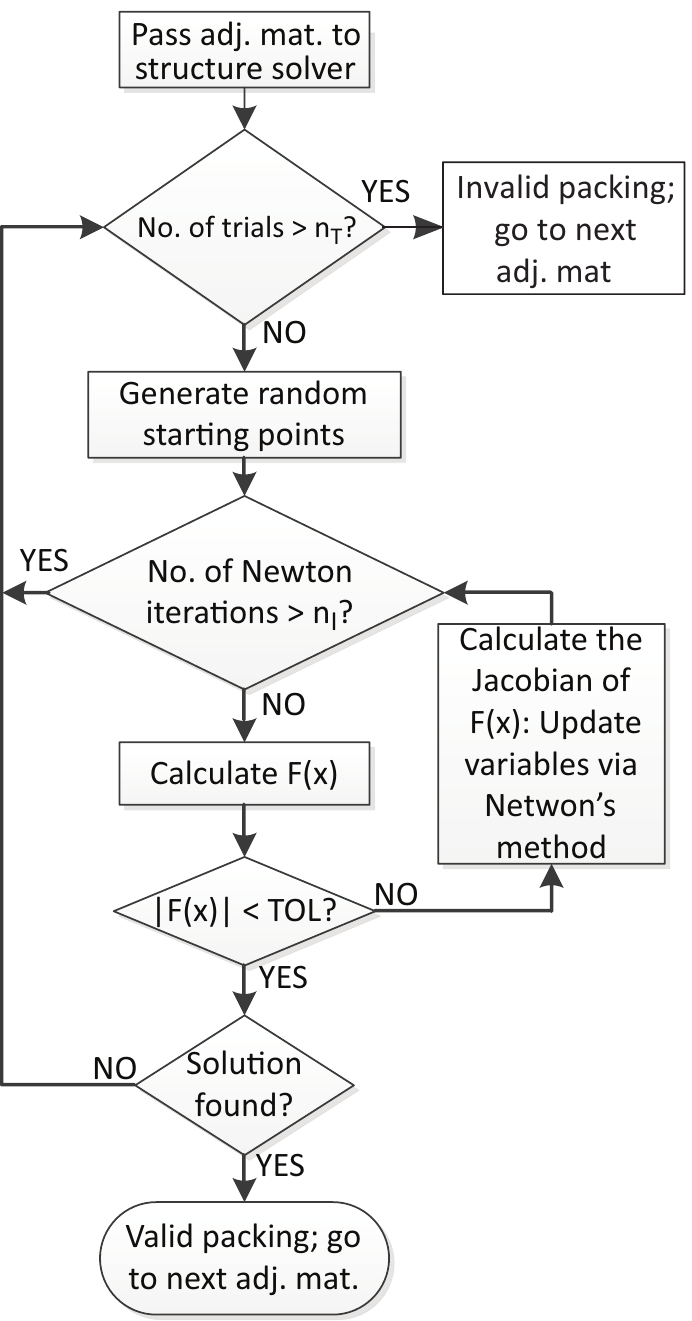}
\caption{Schematic of Part 2 of our enumeration protocol; Euclidean structure solver.  This diagram corresponds to the third box in Fig.\ \ref{fig:topschem}.  All symbols ($n_T,\ n_I,\ F(x),\ TOL$) are described in the text.}
\label{fig:solverschematic}
\end{figure}

We solve for the Euclidean structure $\{\vec{r}\}$ of nonisomorphic packings which are not eliminated by any of the geometric rejection rules.
The adjacency matrix yields a set of $N_c$ equations and $N(N-1)/2 - N_c$ inequalities for $i \in [1,N], j \in [i+1,N]$:
\begin{equation}
\begin{array}{ccccc}
|\vec{r}_i - \vec{r}_{j}|^2 & = 1 & ; & A_{ij} = 1\\
& & & & \\
|\vec{r}_i - \vec{r}_{j}|^2 & \geq 1 & ; & A_{ij} = 0.
\end{array}
\label{eq:cartesianset}
\end{equation}
Solutions to Eq.\ \ref{eq:cartesianset} are valid $N/N_c$ packings except in the case where they possess
``implicit'' contacts corresponding to the ``$=$'' case of the ``$\geq$''.
Proper accounting of implicit contact graphs is key to exact enumeration studies, both for determining $\mathcal{M}$ and for developing graph-theoretic rejection rules; see Appendix \ref{sec:implic} for a discussion of these issues.

We solve Eq.\ \ref{eq:cartesianset} efficiently using a multidimensional Newton solver with step size control \cite{dahlquist74} schematically depicted in Fig.\ \ref{fig:solverschematic}.
Initial conditions for the solver $\{\vec{r}_{init}\}$ are generated by placing $N$ particles randomly within a cube of length $N$, centered at the origin. 
The solver then attempts to find the roots of 
\begin{equation}
\begin{array}{c}
F(\vec{r}) =  \sum_{j>i} \delta \left(A_{ij} - 1\right) \left( \left| \vec{r_i} - \vec{r_j} \right|^2-1 \right) + \\
R \sum_{j>i} \delta \left(A_{ij} \right) \left( \left| \vec{r_i} - \vec{r_j} \right|^2-1 \right) \Theta\left(1-\left| \vec{r_i} - \vec{r_j} \right|\right) =  0,
\end{array}
\label{eq:Fofvecr}
\end{equation}
where $F(\{\vec{r}\})$ is the ``error'' function, $\delta$ is the Kronecker delta function, and $\Theta$ is the Heaviside step function with $\Theta(x) = 0$ for $x \leq 0$ and $1$ for $x > 0$.
The first term in Eq. \ref{eq:Fofvecr} enforces non-overlapping contact between particle pairs with $A_{ij} = 1$ and the second term is a repulsive term penalizing overlaps for particles with $A_{ij} = 0$.
For $R = 10$ the combination of repulsive force and step size control gives a large (order-of-magnitude) speedup over a version lacking these features.

The iterative nature of our solver is illustrated in Fig.\ \ref{fig:solverschematic}.
Solutions are considered converged and a valid packing is found when $|F(\{\vec{r}\})| < TOL$. 
If $|F(\{\vec{r}\})| > TOL$ after $n_I$ Newton iterations, the solution is discarded and the process begins with a new  $\{\vec{r}_{init}\}$.
If a solution is not found after $n_T$ attempts, $\bar{A}$ is rejected as an invalid packing.
We find that failure of convergence of the structure solver to converge to $|F(\{\vec{r}\})| < TOL$ within $n_T$ attempts is sufficient to reject invalid packings, provided $n_T$ is sufficiently large.
This sampling over different  $\{\vec{r}_{init}\}$ is an important part of our enumeration procedure since the set of geometric rejection rules remains incomplete for $N > 9$; further details are given in Appendix \ref{sec:solvervalidation}.

\subsection{Dynamical matrix analyses}
\label{subsec:dynmat}

Determining mechanical stability of packings is of great interest since packings with ``floppy'' modes have high vibrational entropy.
We determine stability using dynamical matrix analyses \cite{jacobs95}.  
The Hessian matrix 
\begin{equation}
\displaystyle\frac{\partial^2 U}{\partial\vec{r}_{i}\partial\vec{r}_{j}},
\label{eq:dynmat}
\end{equation}
has $3N-6$ positive eigenvalues for mechanically stable packings, but fewer for floppy packings.
Since Eq.\ \ref{eq:stickyspherepot} is singular at $r = 1$, we (following Ref.\ \cite{hoy10}) replace it by 
\begin{equation} 
U_{harm}(r) = \bigg{\{}\begin{array}{ccc}
-\epsilon + \displaystyle\frac{k_c}{2}\left(r-1\right)^{2} & , & r < r_c\\
& &\\
0 & , & r > r_c
\end{array}
\label{eq:perturbed}
\end{equation}
with $U = \sum_{j > i} U_{harm}(r_{ij})$.  Note that Eq.\ \ref{eq:perturbed} reduces exactly to the sticky sphere potential (Eq.\ \ref{eq:stickyspherepot}) in the limit $k_c \to \infty$.
We choose $k_c = 10^{5}\epsilon$ and $r_c/D = 1 + \sqrt{2\epsilon/k}$.
For this $k_c$, in all cases (since implicit-contact graphs are eliminated; see Appendix \ref{sec:implic}), only the  $N_c$ pairs specified by $\bar{A}$ interact via $U_{harm}$.

\subsection{Structural Analyses}
\label{subsec:structanaly}

We analyze the structural order of nuclei using several measures: crystallographic point group symmetry, compatibility with the bulk crystal, and the presence of various structural motifs within nuclei.

Point groups provide a convenient means of classifying sticky sphere packings.  
Macrostates with higher symmetry have lower permutational entropy \cite{meng10} and are often associated with higher crystalline order.
We evaluate point group symmetries of packings using the Euclidean solutions for $\{\vec{r}\}$ and the symmetry evaluator of Lee and Shattuck \cite{shattuck}.

Barlow packings \cite{barlow1883,barlowfoot} are hard-sphere packings composed of layered hexagonal-close-packed planes; their three-dimensional order may be FCC, HCP, or mixed FCC/HCP, but they possess no defects (\textit{e.g.}\ stack faults.)  
They are optimal nuclei for hard- and sticky-hard-sphere crystals since they possess the same ordering as the bulk equilibrium crystals, and have $\phi \to \pi/\sqrt{18}$ in the $N \to \infty$ limit.
We identify nuclei with Barlow order by verifying that all $\{r_{ij}^2\}$ are equal to values found in Barlow packings, \textit{i.e.}\ $\{r_{ij}^2\} \in \{0, 1, 2, 8/3, 3, 11/3, 4, 5, 17/3, 6, 19/3, 20/3, 7, 22/3, 8,\\ 25/3, 9, 29/3, 10, 31/3, 32/3, 11, 34/3, ...\}$ for all $i$ and $j$.

Structural motifs relevant to nucleation, \textit{e.g.}\ 5- and near-5-fold symmetric structures and stacking faults, are identified through the presence of  $M\times M$ subgraphs $\bar{A}'$ uniquely associated (Section \ref{subsec:euclidean}) with the corresponding structures within the $N \times N$ adjacency matrices.
For motifs associated with a pattern $X$, we identify the number of macrostates $\mathcal{M}_{X}$ and fraction of microstates $f_{X}$ including these patterns via
\begin{equation}
\mathcal{M}_{X}(N,N_c) = \sum_{k=1}^{\mathcal{M}(N,N_c)} G(X),
\label{eq:patternM}
\end{equation}
where $G_k(X)$ is $1$ if structure of the $k^{th}$ macrostate matches the pattern and $0$ otherwise, and
\begin{equation}
f_{X}(N,N_c) = \Omega^{-1} \sum_{k=1}^{\mathcal{M}(N,N_c)} \omega_k G_k(X),
\label{eq:patternf}
\end{equation}
where $\{\omega\}$ and $\Omega$ are given by Eqs.\ \ref{eq:omegak}-\ref{eq:omega}.

We choose to identify these motifs as described above rather than alternatives such as determining the number of ``crystal-like'' particles as is common practice in the literature \cite{tenwolde96}. 
The latter practice is better suited to studies of bulk crystallization, whereas we consider small nuclei where surface effects dominate. 
For example, while many studies have examined formation of nuclei with Barlow (FCC, HCP, and RHCP) ordering, we are not aware of any previous studies that quantitatively examined the \textit{fraction} $f_{Barlow}$ of nuclei possessing such order.

\section{Results}
\label{sec:results}

We now report results for the number, structure, symmetry, and dominant structural motifs within all isostatic and hyperstatic sticky sphere packings for $N \leq 11$.
Results for $\mathcal{M}(N,N_c)$ are shown in Table \ref{tab:summarytab}.
Values for $N \leq 10$ are the same as reported in Refs.\ \cite{hoy10,arkus11}.
$\mathcal{M}$ grows exponentially with increasing $N$.
Exponential growth is expected for systems with short-range interactions and liquidlike order \cite{stillinger82}, but has not previously been conclusively demonstrated for sphere packings \cite{footgao}.
The arguments by Stillinger and Weber in Ref.\  \cite{stillinger82} apply to ``large'' N; our results suggest that sticky sphere packings are already in the large-$N$ limit for $N \geq 9$.

To examine the degree of crystalline versus liquidlike order, we report the number of macrostates with $C_1$ point group ordering and Barlow ordering ($\mathcal{M}_{C_1}$ and $\mathcal{M}_{Barlow}$ respectively).
We also report the fraction of microstates $f_{C_1}$ and $f_{Barlow}$ with $C_1$ and Barlow order.
%these respective orderings. 
For isostatic packings, $f_{C_1}$ increases rapidly with $N$, while $f_{Barlow}$ decreases rapidly for $N \geq 7$.
Fractions of nuclei with Barlow order increase sharply with hyperstaticity, consistent with the onset \cite{torquato00} of crystallization for $N_c > N_{ISO}$.

The arguments of Phillips and Thorpe \cite{phillips79} that glass-formation is optimized when systems are isostatic have been supported by many studies, including studies of systems interacting via central forces.
Our results support these arguments. 
$\Omega$ decreases sharply with increasing hyperstaticity (e.g.\ $\Omega(N= 11,\ N_c = 29)/\Omega(N= 11,\ N_c = 27) =  8.56\cdot10^{-5}$), indicating an entropic barrier to increasing $N_c$ beyond $N_{ISO}$.
A large fraction of isostatic nuclei have liquid-like symmetry yet are solidlike in character (\textit{i.e.} mechanically stable \cite{footfloppy} -see Appendix \ref{sec:bridge}), and cannot change structure without breaking bonds, indicating (for sticky spheres) an enegetic barrier to increasing $N_c$ beyond $N_{ISO}$.
These results provide a quantitative (if partial) explanation for earlier reports of glass formation by kinetic arrest in sticky hard sphere systems \cite{foffi00,buzzacaro07,royall08}.
Specifically, both energetic and entropic barriers should impede nucleation of more ordered crystallites; our quantification of these effects may help explain why classical nucleation theory breaks down for $N \lesssim 10$ \cite{auer01}.
In the following subsections, we will examine the shapes, symmetries, and relevant structural motifs of sphere packings with $N \leq 11$ in quantitative detail.
We found 99\% of these packings to be mechanically stable; results of our stability analyses are discussed in Appendix \ref{sec:bridge} \cite{footnobridge}.

\begin{table}[htbp]
\caption{Numbers of macrostates $\mathcal{M}$, macrostates with liquid-like ($C_1$) symmetry $\mathcal{M}_{C_1}$, macrostates with Barlow ordering $\mathcal{M}_{Barlow}$, and fractions of microstates with $C_1$ symmetry and Barlow ordering ($f_{C1}$ and $f_{Barlow}$, respectively) \cite{foottech}.  Results for $\mathcal{M}$ for $N \leq 10$ are the same as reported in Ref.\ \cite{arkus11}. Note that some Barlow-ordered nuclei can have $C_1$ symmetry, \textit{i.e.}\ $f_{Barl} + f_{C1} > 1$.  $^*$: $\mathcal{M}(11,27)$ excludes the ``bridge'' packings described in Appendix \ref{sec:bridge}.}
\begin{ruledtabular}
\begin{tabular}{lcccccc} 
$N$ & $N_c$ & $\mathcal{M}$ & $\mathcal{M}_{C_1}$ &  $\mathcal{M}_{Barlow}$ & $f_{C_1}$ & $f_{Barlow}$\\
5 & 9 & 1 & 0 & 1 & 0 & 1\\
6 & 12 & 2 & 0 & 1 & 0 & .077\\
7 & 15 & 5 & 0 & 1 & 0 & .612\\
8 & 18 & 13 & 2 & 4 & .089 & .268\\
9 & 21 & 52 & 21 & 11 & .717 & .154\\
10 & 24 & 259 & 188 & 33 & .912 & .115\\
10 & 25 & 3 & 0 & 3 & 0 & 1\\
11 & 27 & 1620$^*$ & 1394 & 103 & .954 &.056\\
11 & 28 & 20 & 8 & 12 & .744 & .488\\
11 & 29 & 1 & 0 & 1 & 0 & 1
\end{tabular}
\end{ruledtabular}
\label{tab:summarytab}
\end{table}

\subsection{``Bulk'' measures of shape and symmetry}
\label{subsec:bulk}

\begin{figure}[htbp]
\includegraphics[width=3in]{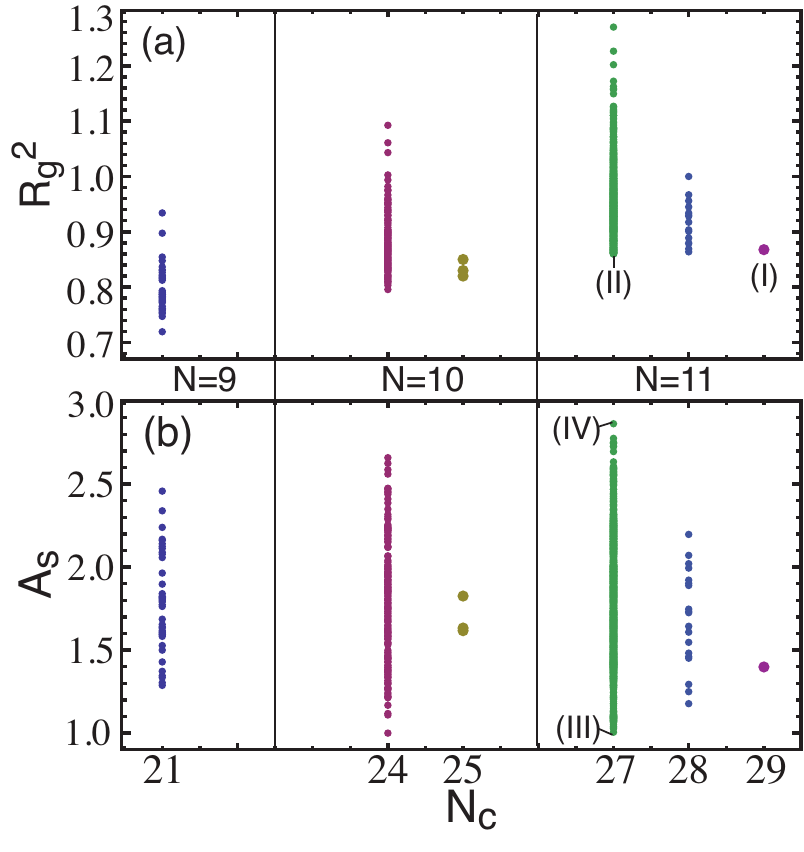}
\caption{(Color online) Distributions of (a) $R_g^2$ and (b) shape anisotropy $\mathcal{A}_s$.  Results from left to right indicate increasing $N$ and $N_c$. Roman numerals (I)-(IV) indicate the extremal packings shown in Fig.\ \ref{fig:mostleast}.}
\label{fig:Rgsq}
\end{figure}

The complete set of packings reported here \cite{onlinepackings} exhibits a great diversity of symmetries and shapes.
Figure \ref{fig:Rgsq}(a) shows values of $R_g^2$, 
\begin{equation}
R_g^{2} = \displaystyle\frac{1}{N} \sum_{i=1}^{N} \left|\vec{r}_{i} - \left< \vec{r}\right> \right|^2,
\label{eq:rgsqformula}
\end{equation} 
for all $9 \leq N \leq 11$ macrostates.
Each data point shows results for one macrostate. 
Two notable features are apparent.
First, for isostatic states, the widths of the distributions $\Delta R_g^{2}/\left<R_g^2\right>$ increase with increasing $N$.
Second, the most compact packings for $N = 10$ and $11$ are \textit{not} the maximally contacting packings.  For example, 25 of the 259 $N = 10$, $N_c = 24$ packings have a smaller $R_g^2$ than the most compact of the 3 $N = 10$, $N_c = 25$ packings, while for $N=11$, the most compact $N_c = 28$ packing and the 66 most compact $N_c = 27$ packings have $R_g^2$ below that of the $N_c = 29$ minimal energy packing.

Figure \ref{fig:Rgsq}(b) shows the shape anisotropy $\mathcal{A}_s$ for the same set of macrostates, \textit{i.\ e.}\ $\mathcal{A}_s(N, N_c) = \sqrt{\lambda_{max}/\lambda_{min}}$ where $\lambda_{max}$  are the maximum and minimum eigenvalues of the moment of inertia tensor $\bar{R}^{2}$:
\begin{equation}
\bar{R}^{2} = \displaystyle\frac{1}{N} \sum_{i=1}^{N} |\vec{r}_{i}\cdot\hat{e}_j - \left< \vec{r}\right>\cdot\hat{e}_k |^2.
\label{eq:rgsqformula2}
\end{equation}
Here $\hat{e}_l$ is the unit vector along the $l$-axis, where $l = x,\ y, \rm{and}\ z$.
Maximally symmetric (sphere-like) packings have $A_s \simeq 1$.
Isostatic packings show a broad range of anisotropy that increases with increasing $N$.  
Anisotropy does not systematically increase with the degree of hyperstaticity. 
However, it is clear that the \textit{range} of anisotropy decreases.  
Both the most and least symmetric packings are isostatic.

To check whether the results in Fig.\ \ref{fig:Rgsq} are representative of the full ensemble of packings, we examine the probability distributions $P(R_g^2)$ and $P(A_s)$, where
\begin{equation}
P(R_g^2)(N,N_c) = \Omega^{-1} \sum_{k=1}^{\mathcal{M}(N,N_c)} \omega_k R_{g,k}^2,
\label{eq:PRg}
\end{equation}
and
\begin{equation}
P(A_s)(N,N_c) = \Omega^{-1} \sum_{k=1}^{\mathcal{M}(N,N_c)} \omega_k A_{s,k},
\label{eq:PAs}
\end{equation}
where $R_{g,k}^2$, $A_{s,k}$, and $\omega_k$ are the squared radius of gyration, anisotropy, and permutational entropy (Eq.\ \ref{eq:omegak}) of the $k^{th}$ macrostate.
  
\begin{figure}[htbp]
\includegraphics[width=3in]{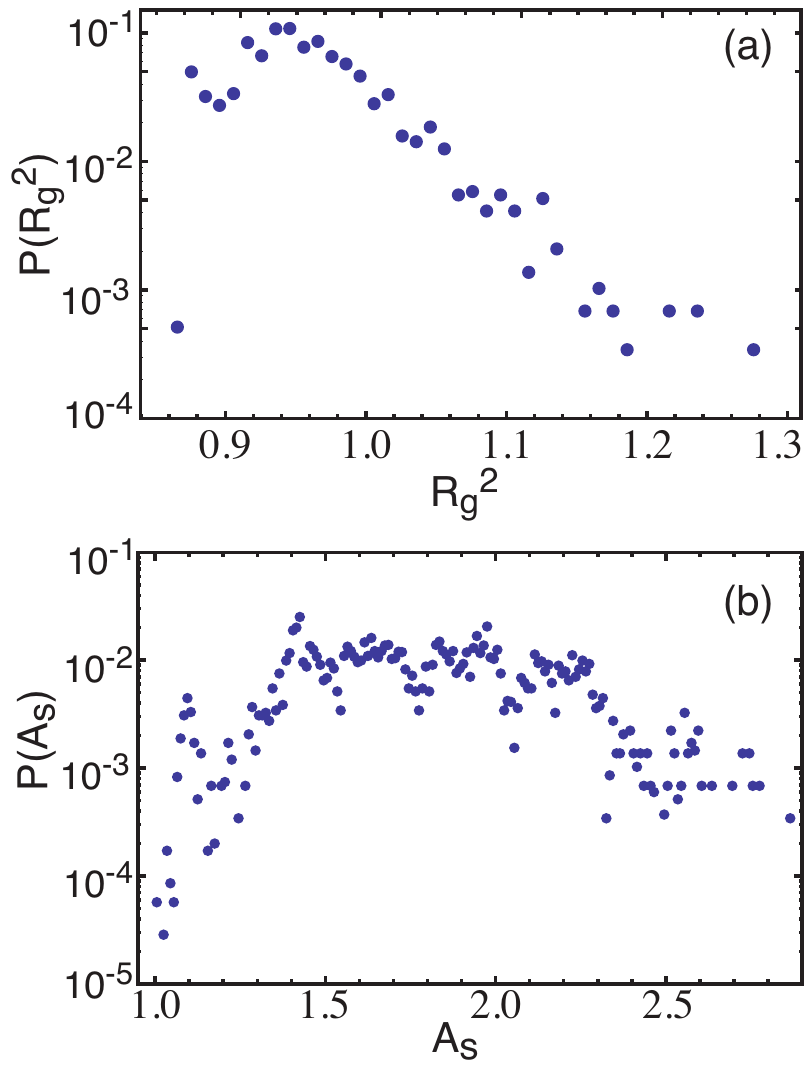}
\caption{(Color online) Probability distributions of (a) $R_g^2$ and (b) shape anisotropy $\mathcal{A}_s$ for $N=11$, $N_c = 27$ packings.}
\label{fig:RgAdistribs}
\end{figure}

Figure \ref{fig:RgAdistribs} shows $P(R_g^2)$ and $P(A_s)$ for $N= 11,\ N_c = 27$.  
Slightly narrower distributions are obtained for smaller $N$, but are qualitatively similar.
Results indicate that the most compact and the most symmetric nuclei have low entropy and are consistent with earlier studies for smaller $N$ \cite{arkus09,meng10}.
$P(A_s)$ is particularly broad.
These distributions indicate that the ``typical'' nuclei is neither spherical nor characterized by a single value of $R_g$.

\begin{figure}
\includegraphics[width=3.1in]{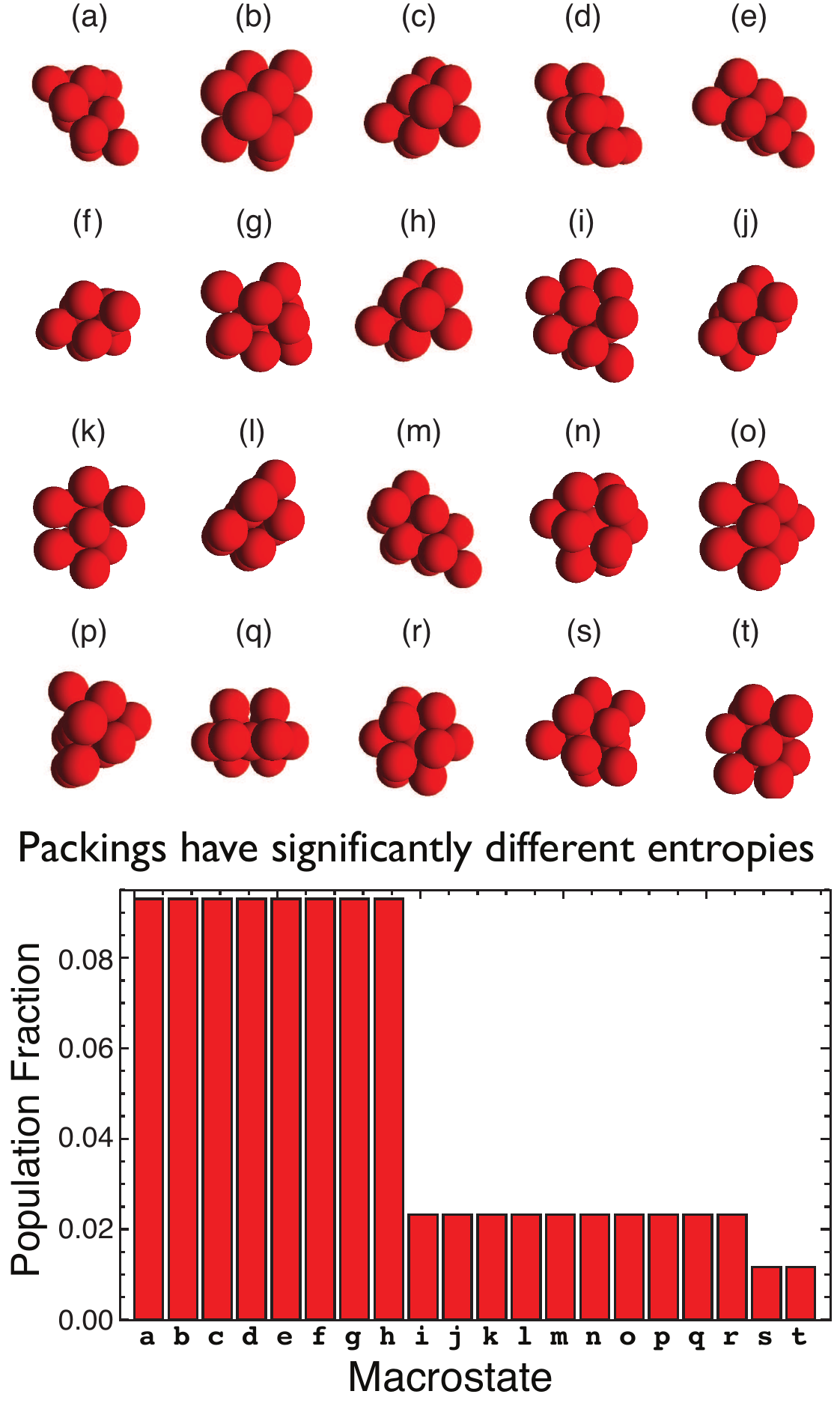}
\caption{(Color online) Top panel: (a-t) The 20 distinguishable macrostates for $N=11,\ N_c = 28$. Bottom panel: Entropic fractions $\omega_k/\Omega$ for these macrostates.  Relative values of $\omega_k$ are inversely proportional the symmetry numbers $c$ of the associated point groups \cite{meng10}, \textit{i.e.}\ the $C_1$ macrostates (a-k) have $c=1$ while the most symmetric macrostates have $c=8$.}
\label{fig:1128}
\end{figure}

Although widths of the distributions of shapes, sizes, symmetries, and entropies decrease for hyperstatic nuclei, they remain broad.
The top panel of Fig.\ \ref{fig:1128} shows the 20 macrostates for $N = 11,\ N_c = 28$; $R_g^2$ and $A_s$ are the same as in Fig.\ \ref{fig:Rgsq}.
The bottom panel shows their relative permutational entropies $\omega_k/\Omega$ (Eqs.\ \ref{eq:omegak}-\ref{eq:omega}).

The structural diversity of packings reported above illustrates a key feature that should be included in theoretical treatments of nucleation in hard- and sticky-hard sphere systems.  
Figure \ref{fig:1128} shows why one would not expect classical nucleation theory to work for sticky hard sphere packings in this $N$-regime and it is neccessary to consider nuclei possessing arbitrary geometry.
There has been great interest in recent years in finding the densest finite sphere packings \cite{torquato10, hopkins11}, \textit{i.e.}\ the $N$-sphere nuclei that minimize volume $V$.
Most studies (\textit{e.g.}\ Refs.\ \cite{hopkins11,wales10,torquato10}) search for nuclei optimizing either density or energy.
For hard- and sticky-hard spheres, however, it is far from clear which quantity one should optimize.
We have shown in detail that the relation between the number of contacts and density is nontrivial, and that packings optimizing these two features are in general different from each other.
While this competition will break down at large $N$, \textit{i.e.}\ the FCC crystal is simultaneously the densest packing and a maximally contacting packing for $N \to \infty$ \cite{hales98}, and the $N^{*}$ at which the crossover occurs is unknown, we have shown that $N^{*} > 11$ \cite{footclassicalN}.

\begin{figure}[htbp]
\includegraphics[width=3.25in]{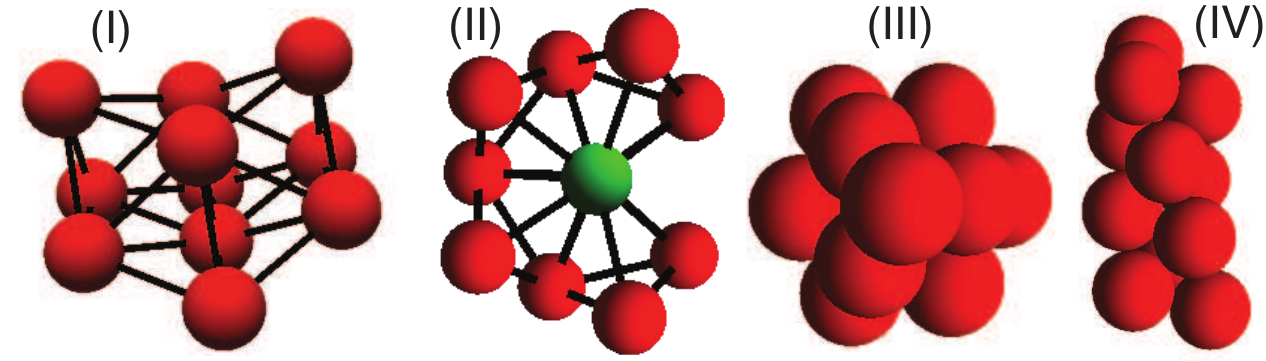}
\caption{(Color online) The (I) ground, (II) most spherically compact, (III) densest,and most symmetric, and (IV) least compact and symmetric packings for $N = 11$.  Packings (II-IV) are all isostatic ($N_c = 27$) and therefore second excited states.  Sphere radii in panels (I-II) have been reduced for clarity.}
\label{fig:mostleast}
\end{figure}

Figure \ref{fig:mostleast} further illustrates the competition between energy minimization and density maximization, and the importance of asphericity therein, by contrasting ``extremal'' packings for $N = 11$.
The unique sticky hard sphere ground state (panel (I)) has $N_c = 29$ and HCP order \cite{arkus09} but is neither the densest, most compact, nor most symmetric.
Panel (II) shows the packing that fits within the smallest \textit{spherical} volume.
It would be an ideal nucleus for hard-sphere crystallization within the framework of classical nucleation theory, but is in fact a ``bad'' (off-pathway) nucleus since it is a partial icosohedron lacking Barlow order.
Panel (III) shows the \textit{densest} packing (in the sense that it fits within the smallest convex shrink-wrapped volume $V_{sw}$), that has FCC order.
It is also the most symmetric packing, \textit{i.e.}\ the nucleus that minimizes $A_s$.
Finally, panel (IV) shows the packing that is simultaneously the least compact and least spherically symmetric; it is also a ``bad'' nucleus lacking Barlow order. 
Note that the packings shown in panels (II-IV) are all second excited states for sticky spheres, energetically degenerate, and (except for (IV)) mechanically stable.

\subsection{Nuclei with structural motifs incompatible with bulk crystallization}
\label{subsec:hostilemotifs}

Stacking faults and five-fold symmetric structures are ``defects''
incompatible with bulk crystallization at $\phi = \pi/\sqrt{18}$.
Their presence is well-known to impede hard-sphere crystallization
\cite{buzzacaro07, omalley03,zaccarrelli09,karayiannis11}.  In this
subsection, we quantify the propensity of nuclei to contain these and
related structural motifs.

The simplest stack-faulted motif is the  $M = 6,\ M_c = 12$ capped trigonal bipyramid structure shown in Fig.\ \ref{fig:macrostateexamp}(b).
Table \ref{tab:hostilemotifs} shows the number of macrostates $\mathcal{M}_{ctb}$ and fraction of microstates $f_{ctb}$ that include this motif.
Note that $\mathcal{M}_{ctb}$ and $f_{ctb}$ are \textit{lower bounds} for the numbers and fractions of stack-faulted structures since other stack-faulted motifs exist.
Nonetheless, the propensity for stack-faulting is surprisingly high given the small size of the nuclei - above 50\% for all $N > 8$ packings with $N_c < N_c^{max}$.
Stack faults appear in hyperstatic nuclei at $N=11$, which is consistent with the fact that these nuclei do not maximize contacts.

\begin{figure}[htbp]
\includegraphics[width=2.5in]{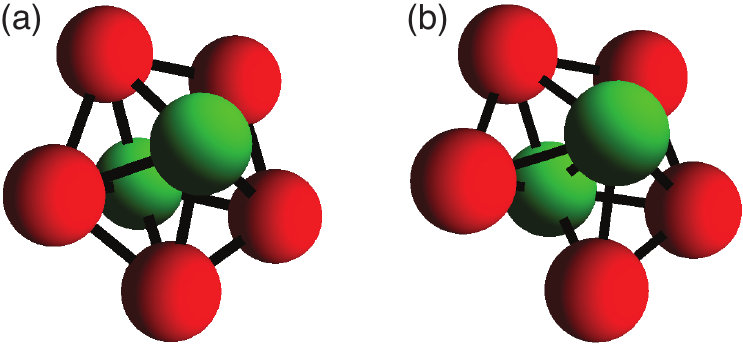}
\caption{(Color online) Some structural motifs hostile to bulk crystallization: (a) The $M = 7$, $M_c = 15$ partial icosohedral and (b) $M = 7$, $M_c = 15$ open-loop near-5-fold symmetric structures.}
\label{fig:picosop}
\end{figure}

Two five-fold-like structural motifs often found in small nuclei are seven-sphere minimal energy packings ($M_c = 15$) and are shown in Fig.\ \ref{fig:picosop}.  
Panel (a) shows a five-fold-symmetric partial icosohedron.
Panel (b) shows an \textit{nearly} 5-fold symmetric structure which differs from (a) in that the green (lightest shaded) dimers contact and the 5-ring is open rather than closed with a separation $r_{ij} = 1.19$  \cite{arkus09}.
Both of these structures are incompatible with close-packed crystal structure.

\begin{table*}[htbp]
\caption{Structural motifs that are incompatible with bulk crystallization as a function of $N$ and $N_c$. \textit{Left columns}: Propensity of nuclei to contain minimal stacking faults.  Note that $f_{ctb}$ is strictly $\geq f_{5b}$ since the stack-faulted structure shown in Fig.\ \ref{fig:macrostateexamp}(b) 
is a subset of the near-five-fold symmetric structure shown in Fig.\ \ref{fig:picosop}(b). 
\textit{Middle columns}:  
Number of macrostates and percent of microstates containing partial-iscohedral five-fold symmetric ($\mathcal{M}_{5a}$ and $f_{5a}$) and open-loop ($\mathcal{M}_{5b}$ and $f_{5b}$) near-5-fold symmetric substructures in Fig.\ \ref{fig:picosop}.  
\textit{Right columns}: Propensity of nuclei to contain second nearest neighbors with $r^{min}_{2nd} < \sqrt{2}$; values of $r^{min}_{2nd}$, numbers of macrostates $\mathcal{M}_{<\sqrt{2}}$ and fractions of microstates $f_{<\sqrt{2}}$ with $r^{min}_{2nd} < \sqrt{2}$.  
Values of $r^{min}_{2nd}$ for $N = 5$, $6$, $7$ and $8$ are the same as in Ref.\ \cite{arkus09}.  Note that there are several ``near-miss'' $N = 11$, $N_c = 27$ macrostates with $r^{min}_{2nd}$ only slightly above 1.  Some of these possess soft modes, but we have verified that these cannot form 28th contacts.}  
\begin{ruledtabular}
\begin{tabular}{lcccccccccc}
$N$ & $N_c$ & $\mathcal{M}_{ctb}$ & $f_{ctb}$ & $\mathcal{M}_{5a}$ & $f_{5a}$ & $\mathcal{M}_{5b}$ & $f_{5b}$ & $r^{min}_{2nd}$ & $\mathcal{M}_{<\sqrt{2}}$ & $f_{<\sqrt{2}}$\\
7 & 15 & 3 & .286 &1 & .102 & 1 & .153 &  $\sqrt{2 - 2/\sqrt{5}}$ & 2 & .255\\
8 & 18 & 7 & .567 & 1 & .153 & 3 & .166 & $4\sqrt{6}/9$ & 5 & .331\\
9 & 21 & 30 & .700 & 7 & .098 & 22 & .512 &  $\sqrt{2 - 2/\sqrt{5}}$ & 33 & .691\\
10 & 24 & 165 & .643 & 32 & .135 & 110 & .415 & 1.03296 & 185 & .721\\
10 & 25 & 0 & 0  & 0 & 0 & 0 & 0  & $\sqrt{2}$ & 0 & 0\\
11 & 27 & 1126 & .723 & 220 & .130 & 726 & .467 &  1.00489 & 1332 & .835\\
11 & 28 & 8 & .511 &  0 &  0 & 1 & .0116 & $4\sqrt{6}/9$ & 2 & .035\\
11 & 29 & 0 & 0 & 0 & 0 &  0 & 0 & $\sqrt{2}$ & 0 & 0 
\end{tabular}
\end{ruledtabular}
\label{tab:hostilemotifs}
\end{table*}

Table \ref{tab:hostilemotifs} shows the fraction of macrostates for $7 \leq N \leq 11$ containing motifs that are incompatible with long-range crystalline order.
For isostatic packings, the 5-fold symmetric subclusters in Fig.\ \ref{fig:picosop} are found in many macrostates and about 60\% of microstates for $9 \leq N \leq 11$.
In contract, these 5-fold symmetric subclusters are found in only one of the 24 nonisomorphic hyperstatic packings (and in $\lesssim 1\%$ of microstates).
The near-5-fold symmetric structure shown in Fig.\ \ref{fig:picosop}(b) corresponds to an elementary \textit{twin} defect.
The high fraction of these ($f_{5b}$) may explain why five-fold symmetric twinned crystallites are commonly observed in sticky- and hard-sphere systems \cite{omalley03,buzzacaro07}.

Another metric for nuclei incompatible with bulk crystallization at $\phi = \pi/\sqrt{18}$ is the minimum 2nd-nearest neighbor distance $r^{min}_{2nd}$.
FCC-, HCP-, and Barlow-ordered crystallites have $r^{min}_{2nd} = \sqrt{2}$.
Therefore nuclei with $N > 6$ and $r^{min}_{2nd} < \sqrt{2}$ cannot have Barlow order \cite{footoct}.
Table \ref{tab:hostilemotifs} shows the numbers of macrostates $\mathcal{M}_{<\sqrt{2}}$ and fractions of microstates $f_{<\sqrt{2}}$ with $r^{min}_{2nd} < \sqrt{2}$.  
The fact that $f_{<\sqrt{2}} + f_{Barlow} < 1$ for all $N$ and $N_c$ does not indicate any inconsistency since
stack-faulted structures tend to be associated with neighbor distances $r_{ij} > \sqrt{2}$,

In the above subsections, we have examined several measures of microstructural order.  $f_{Barlow}$ is a measure of ``good'' nuclei that are consistent with long-range crystalline order (LRCO) while $1-f_{Barlow}$, $f_{5a} + f_{5b}$, $f_{ctb}$, and $f_{<\sqrt{2}}$ are four independent measures of ``bad'' nuclei that are inconsistent  with LRCO \cite{foottech}.
The latter four all show the same trends; they increase with $N$ for isostatic packings (to very high fractions) and decrease with increasing hyperstaticity.  
High energy barriers are expected between ``bad'' and Barlow-ordered nuclei since many bonds must be rearranged to change from one ordering to the other.  
These results provide an explanation for the propensity of sticky hard sphere systems to jam and glass-form in both simulations and experiments.

\subsection{Applicability to other potentials and methods}
\label{subsec:applic}

Our results for sticky hard-sphere packings are relevant for analyses
of clusters formed by systems that interact via other potentials
$\mathcal{U}(r)$ with hard-core like repulsions and short range
attractions.  The sticky hard sphere model has been shown to provide a
perturbative ``reference state'' \cite{cochran06} for such potentials
(the sphere diameter $D$ may be replaced by the minimum of a general
interparticle potential).  In other words, packings of sticky spheres
become increasingly similar to equilibrium clusters of particles
interacting via a potential $\mathcal{U}(r)$ as the range of
interaction $r_c$ (\textit{i.\ e.} $\mathcal{U}(r) = 0$ for $r > r_c$)
approaches $D$ from above, and are rigorously identical in the limit
of hard core repulsions and $r_c/D \to 1$.

The minimum 2nd-nearest neighbor distance $r^{min}_{2nd}$ (Table \ref{tab:hostilemotifs}) is a particularly useful metric for evaluating the sticky hard sphere model's suitability for determining minimal energy clusters of other potentials since additional local free energy minima begin to appear when $r_c \gtrsim 1 + (r_{min}^{2nd}-1)/2$ \cite{wales10}.
Our complete set of sticky hard sphere packings form an arguably complete set of initial guesses for identifying all isostatic clusters of up to 9 particles when the interaction range $(r_c/D -1) \lesssim .025$, and for up to 10 particles when the interaction range $(r_c/D -1) \lesssim .015$.
The hyperstatic packing sets should be suitable initial guesses for larger $r_c$; for example, $r_c/D \lesssim 1 + (\sqrt{2}-1)/2$ for our minimal energy ($25-$ and $29-$contact) packings for $N = 10$ and $11$, respectively.

For larger $r_c$, ``initial guesses'' for strain-free \cite{wales10}
minimal energy clusters can be obtained by selecting a subset of the
sticky sphere packings satisfying $1 + (r_{min}^{2nd}-1)/2 \lesssim
r_c$ \cite{footneedmin}.  Given a complete set of initial guesses for
the set of nonisomorphic clusters and their permutational entropies
(\textit{i.e.}\ $\{\vec{r}\}$ and $\{\omega\}$ for each of the
$\mathcal{M}(N, N_c)$ nuclei), sophisticated energy-landscape and
transition-state analyses useful in cluster physics may be performed
\cite{wales04,wales10}.  Potentials for which such a procedure should
be applicable include the ``narrow'' square well, the short-range
limit of the Asakura-Oosawa and Morse potentials
\cite{meng10,wales10}, and the hard-core attractive Yukawa potential
in the strong screening limit.  Such potentials describe a broad range
of physical systems ranging from colloids interacting via
depletion-mediated attractions \cite{buzzacaro07,meng10} to buckyballs
\cite{royall11}.  Further, the ground and mechanically stable excited
states of systems interacting via Eq.\ \ref{eq:stickyspherepot} have
been shown to describe the structure of real colloidal crystallites in
dilute solution at $k_B T \simeq 4\epsilon$
\cite{meng10,perry12,footM}.  These results provide justification for
the use of Eq.\ \ref{eq:stickyspherepot} in our exact enumeration
studies.

Additionally, our results should be useful in numerical implementations of modern theories for aspherical nuclei based on cluster expansions (Mayer f-bond diagrams \cite{mayer42}). Such models are commonly used in liquid-state theory \cite{baxter68,stell91,hhgw11}, and are in principle exactly soluble, but have to date suffered from incomplete sets of Mayer diagrams describing differently structured aggregrates as well as ``dangerous'' \cite{stell91} singular cluster integrals that impede implementation of such theories.
The contact graphs (\textit{i.e.}\ valid adjacency matrices) reported here \cite{onlinepackings} correspond to a complete set of Mayer f-bond diagrams for ground state aggregrates of $M \leq 11$ particles as well as stable first excited states (for $M = 10$) and both first and second excited states for $M=11$, while the implicit contact graphs (Appendix \ref{sec:implic}) correspond to cluster integrals that are singular.

\section{Discussion}
\label{sec:discussion}

An advantage of exact enumeration is that it identifies all nuclei
that \textit{can} form, as opposed to only those that \textit{do} form
for a specific preparation protocol.  However, it treats $N$-particle
nuclei in isolation and neglects solvent effects.  Sticky hard-sphere
nuclei in a solution of other sticky hard spheres would be
``continuously bombarded by and grow by absorbing smaller clusters''
\cite{crocker10}.  Such collisions can influence the
pathways by which small nuclei form larger crystallites.  For example,
an ``off-pathway'' (non-Barlow) nucleus might be excited by a collision
and reform into a larger ``on-pathway'' Barlow nucleus.  Additionally,
the detailed structure of nuclei in such a solvent would be altered
both by finite temperature (\textit{e.g.} vibrational entropy
\cite{meng10,perry12}) and the crystallite-fluid interfacial free
energy \cite{mossa03}.

Many Monte Carlo studies have examined crystallization in bulk
hard-sphere systems
\cite{schilling10,omalley03,zaccarrelli09,karayiannis11}.  Schilling
\textit{et.al.}\ \cite{schilling10} performed Monte Carlo simulations
of crystallization in dense ($\phi=0.54$) hard-sphere liquids and
argued that crystallization occurs through a two-step process wherein
(1) dense ``amorphous'' clusters form and act as (2) ``precursors''
for nucleation of larger close-packed crystallites.  They identified
the growth of Barlow order during stage (2) using the bond-orientational order
parameter $q_6$, but did not examine the detailed structure of the
amorphous clusters ({\it i.e.} they did not examine stacking faults or
5-fold symmetric structures) \cite{otherstudiesfoot}.  Our results are
not fundamentally inconsistent with theirs; the non-Barlow-ordered
nuclei we have identified above could correspond to their
``amorphous'' clusters.  We also note that differences between
sticky-hard and purely repulsive hard spheres will significantly alter
the physics of the (1)$\rightarrow$(2) process since non-Barlow
clusters must break bonds to rearrange into Barlow clusters, with a
corresponding energetic cost.

In relating our studies to crystal nucleation we assume that
mechanically stable nuclei play a key role.  This claim is clearly
well supported in the dilute regime where solvent effects are minimal
\cite{meng10,hoy12}.  Additionally, our contention that understanding
the statistical-geometrical properties of small nuclei (and in
particular, the prominence of structural motifs that are incompatible
with the bulk crystal) is relevant to glass-formation and jamming is
consistent with recent experimental work by Royall \textit{et.al.}\
\cite{royall08}, which indicate that such local motifs lead to kinetic
arrest in colloidal suspensions possessing hard-core repulsive and
short-range attractive interactions.  In the semidilute or
concentrated regimes in which solvent effects are stronger, while our
study cannot capture all the complexities of nucleation from the bulk,
quantitative comparison of the nuclear structures reported here to
those reported in studies of crystallizing liquids
\cite{mossa03,miller04,taffs10} is an interesting topic for future
work.\newline

\section{Conclusions}
\label{sec:conclude}

In this manuscript, we described the structural properties of a complete set of isostatic and hyperstatic packings for hard spheres obtained via exact enumeration.
For sticky hard spheres with contact attractions, we also analyzed mechanical stability.
Our key findings included exponential growth in the number of nonisomorphic isostatic packings and nontrivial variation of the size and symmetry of packings with increasing hyperstaticity.
We also calculated the absolute and relative entropies of all packings and their propensity to include various structural motifs that are either compatible or incompatible with bulk crystallization at $\phi = \pi/\sqrt{18}$.
Isotatic nuclei form an increasingly liquid-like ensemble as $N$ increases.
For example, the fraction $f$ of isostatic nuclei possessing Barlow-order decreases rapidly with $N$ to only about 5\% for $N=11$, and the remaining 95\% contain defects such as stacking faults and 5-fold-symmetric substructures.
While $f$ increases with hyperstaticity $H \equiv N_c - N_{ISO}$, $f$ is only about 50\% for $N=11,\ H = 1$ nuclei.
Although we terminated our enumeration studies at $N = 11$ due to the limits of current computational resources, the trends reported here should \cite{omalley03,torquato00} continue to hold for higher $N$.

Additionally, we have shown that considering nuclei with $N \sim 10$
captures a complex regime \cite{footclassicalN} where classical
nucleation theory performs particularly poorly \cite{auer01}.  In this
regime, maximizing density and maximizing $N_c$ compete, the
distributions of nuclear size and symmetry are broad, and many nuclei
are highly aspherical.  Since colloids with hard-core-like repulsions
and short-range attractions form stable nuclei in this size regime
\cite{meng10,footharrowell}, our results present challenges for
traditional theoretical approaches to nucleation in sticky hard sphere
and related systems.  Most analytic and seminumerical treatments
(\textit{e.g.}\ phase field theory and the classical density
functional of Cahn and Hilliard \cite{cahn58}) either assume ordering
consistent with the bulk crystalline phase, or allow for ordering
different than that of the bulk crystal but assume that nuclei are
spherical.  Our results suggest that such restrictions in traditional
methods prevent them from capturing the potential complexity of
small-$N$ crystallite nucleation.  Novel theoretical treatments of
nucleation should consider nuclei of both arbitrary order and
arbitrary geometry.

Finally, while there have been many recent detailed studies of
crystallization in hard-sphere systems, there have been relatively few
\cite{foffi00,tenwolde97,hoy12} theoretical studies of the dynamics of
\textit{sticky} hard sphere crystallization.  The higher relative
entropies of cluster formation for less-ordered nuclei should strongly
affect nonequilibrium behavior.  For example, the large fractions of
small nuclei with $C_1$ (\textit{i.e.}\ liquid-like) symmetry,
fivefold symmetry, stacking faults, and other types of non-Barlow
ordering constitute an effective ``entropic'' barrier to nucleation
and growth of large ordered crystalline domains that should play a key
role in controlling the critical quench rate above which these systems
glass-form.  For sticky spheres, mechanical stability of the
non-Barlow nuclei presents an additional energetic barrier to ordered
crystallite growth.  It would be interesting to compare the ensembles
of nuclei produced in nonequilibrium studies of sticky sphere
aggregation to ``ideally prepared'' ensembles (in which all possible
aggregates are obtained) like those reported in this paper.

\section{Acknowledgements}

Our enumeration code uses the Boost Graph Libary
(http://www.boost.org/).  Work on calculating $\mathcal{M}(10,24)$ was
conducted in collaboration with Natalie Arkus, V.\ N.\ Manoharan and
Michael P.\ Brenner.  We thank V.N.M., A.\ B.\ Hopkins and S.\ S.\
Ashwin for stimulating discussions.  Support from NSF Award No.\
DMR-1006537 is gratefully acknowledged.  This work benefited from the
facilities and staff of the Yale University Faculty of Arts and
Sciences High Performance Computing Center and NSF grant No.\
CNS-0821132 that partially funded acquisition of the computational
facilities.

\begin{appendix}

\section{Mechanical stability: Floppy packings and bridge structures for $N=11$}
\label{sec:bridge}

``Floppy'' packings possessing soft modes are of special interest for nucleation studies at finite $T$ since they possess higher vibrational entropy \cite{meng10}.
Table \ref{tab:floppytab} shows values of the number of macrostates and fractions of macrostates possessing ``nontrivial'' soft modes (i.e.\ soft modes in packings that satisfy the two necessary conditions \textbf{(i-ii)} for mechanical stability).
These soft modes correspond to small collective monomer motions that do not break contacts \cite{footfloppy}.
Interestingly, the number of nontrivially floppy macrostates $\mathcal{M}_{floppy}$ increases faster than $\mathcal{M}$ with increasing $N$ over the range $9 \leq N \leq 11$.
Both this result and the exponential increase in $\mathcal{M}$ (Table \ref{tab:summarytab}) are related to the emergence of hyperstatic packings with $N_c^{max}(10) = 3N-5$ and $N_c^{max}(11) = 3N-4$.
The presence of these hyperstatic states both makes it easier to form isostatic packings (by adding low-coordinated spheres to a hyperstatic packing), and increases the likelihood that such packings will be floppy.
However, the latter effect is small for the range of $N$ considered. 

\begin{table}[htbp]
\caption{Number of macrostates and fraction of microstates possessing nontrivial soft modes.  No $N<9$ packings possess soft modes, and none of the $N\leq 11$ packings possess more than one soft mode.}  
\begin{ruledtabular}
\begin{tabular}{lccc}
$N$ & $N_c$ & $\mathcal{M}_{floppy}$ & $f_{floppy}$\\
9 & 21 & 1 & .00427\\
10 & 24 & 4 & .0194\\
10 & 25 & 0 & 0\\
11 & 27 & 31 & .0136\\
11 & 28 &  1 & .0116\\
11 & 29 & 0 & 0   
\end{tabular}
\end{ruledtabular}
\label{tab:floppytab}
\end{table}

\begin{figure}[htbp]
\includegraphics[width=2.75in]{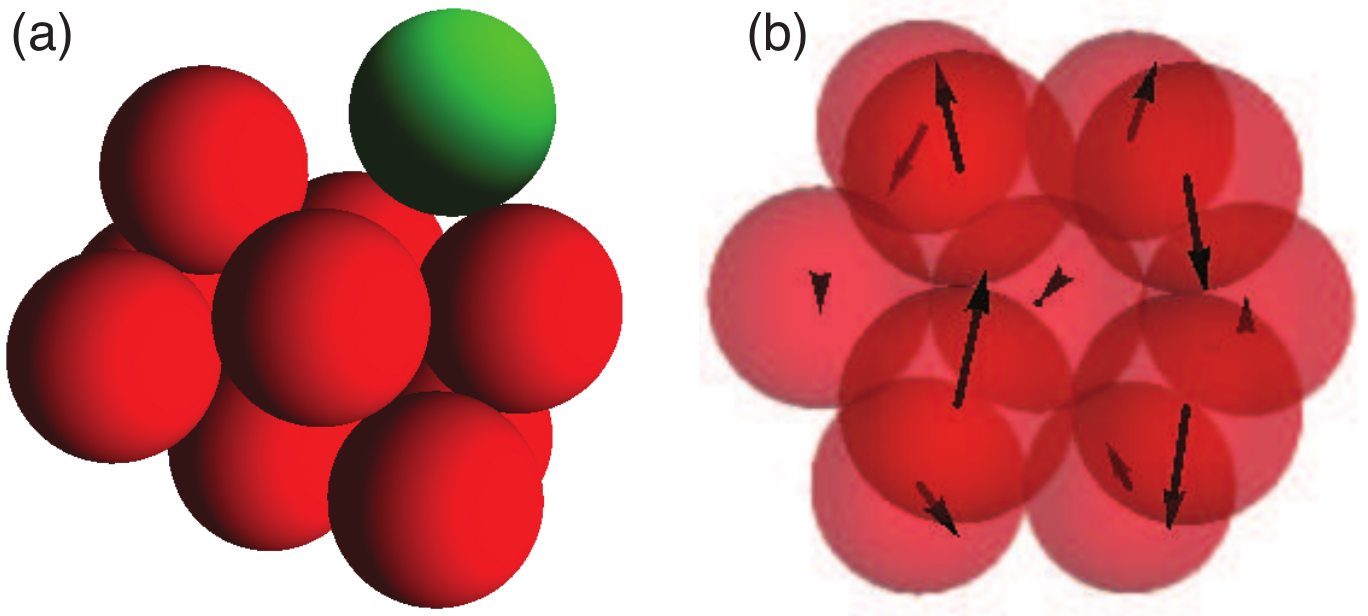}
\caption{Color online) Structure of (a) trivially and (b) hyperstatic floppy nuclei for $N=11$.  Panel (a) shows a typical bridge packing; a bridge monomer possessing only two contacts is shown in green (the lighter shade).   Panel (b): This $N = 11$, $N_c = 28$ packing ((q) in Fig.\ \ref{fig:1128}) is the smallest hyperstatic sticky sphere packing possessing a soft mode.  Arrows indicate particle displacements proportional to a nontrivial zero eigenvector of the dynamical matrix.}
\label{fig:1127bridge}
\end{figure}

Since $10$ is the smallest $N$ at which hyperstatic packings can form, it follows (but has not heretofore been shown) that $11$ is the smallest $N$ at which  ``bridge'' structures can form.
These structures are ``trivially'' floppy because they include monomers that possess only two contacts and fail to satisfy condition \textbf{(i)}. 
We find 25 graph-nonisomorphic $N = 11$, $N_c = 27$ bridge packings.  
A typical example is shown in Fig.\ \ref{fig:1127bridge}(a); bridge spheres (shown in green) have a single floppy mode associated with the free configurational degree of freedom (motion along a circle of $R = \sqrt{3}/2$ centered on the line connecting the two contacted spheres). 
Since we focused on mechanically stable or nontrivially floppy nuclei, these 25 packings are not included in our structural analyses in Section \ref{sec:results}.
However, it is important to include bridge packings in exact enumeration studies that consider finite temperature.
These structures possess high configurational entropy, and for sticky sphere systems, should dominate equilibrium populations of $N = 11, N_c = 27$ nuclei when $k_B T$ is not small compared to the contact energy $\epsilon$.

Another interesting feature of $N=11$ packings is that $11$ is the smallest $N$ at which a hyperstatic sticky sphere cluster with a floppy mode can form.
The packing and associated floppy mode are illustrated in Fig.\ \ref{fig:1127bridge}(b).
This packing has topology similar to a subset of a BCC lattice; 3 adjacent squares of bonds surround a linear trimer.
The floppy mode is a torsional motion about this trimer, and is associated with the ``squares'' of bonds forming the outer part of the packing.

\section{Verification of $\mathcal{M}$ for $N = 11$}
\label{sec:solvervalidation}

\begin{figure}[htbp]
\includegraphics[width=2.75in]{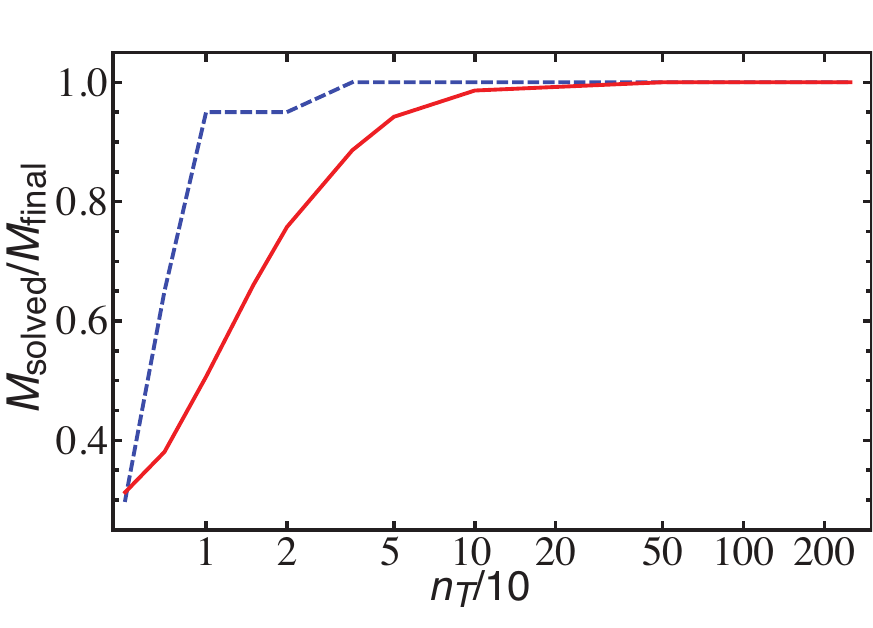}
\caption{(Color online) Convergence of structure-solver for $N = 11$ packings with increasing $n_T$ (see Section \ref{subsec:euclidean}).  $\mathcal{M}_{final} = 20$ for $N_c = 28$ (dotted line) and 1658 for $N_c = 27$ (solid line; this includes the 13 implicit contact packings discussed in Appendix \ref{sec:implic}.).  Here $n_I=40$ and $TOL=10^{-11}$.}
\label{fig:converg}
\end{figure}

For $N > 9$ our set of rejection rules is incomplete and we rely upon the structure solver to determine whether packings are valid.  Here we provide evidence validating this procedure by showing convergence with increasing $n_T$ of the fraction of adjacency matrices identified as valid (see Fig.\ \ref{fig:algschematic}.)
Figure \ref{fig:converg} shows results for a set of 43 nonisomorphic $N = 11$, $N_c = 28$ packings and a set of 4534 nonisomorphic $N = 11$, $N_c = 27$ packings that passed a set of $M \leq 10$ rejection rules from a non-final version of our code.
For fixed $n_T$, the structure solver finds $\mathcal{M}_{solved}$ valid packings.
$\mathcal{M}_{solved}$ converges to $\mathcal{M}_{final}$ in the limit of large $n_T$ and we report $\mathcal{M} = \mathcal{M}_{final}$ in Section \ref{sec:results}.
In all cases, including earlier tests on $N=10$ packings (not shown), convergence is found for $n_T \gtrsim 100$ and increasing $n_T$ by another order of magnitude produces no additional solutions. 
Faster convergence is found for smaller $N$.

\section{Implicit Contact Graphs}
\label{sec:implic}

Equation \ref{eq:cartesianset} is a set of equations and inequalities sufficient to obtain the structure of $N$-sphere packings with \textit{at least} $N_c$ contacts and no overlaps.
The geometric rejection rules enforce only the ``$>$'' portion of the $r_{ij} \geq 1$ condition for $\{i,j\}$ pairs with $A_{ij} = 0$.
For $N \geq 10$, there exist $\bar{A}$ with $N_c$ contacts whose Euclidean solution $\{\vec{r}\}$ is a packing with $N_c + 1$ contacts. 
Such ``implicit contact'' graphs violate the spirit of our enumeration method.
Therefore, all $\bar{A}$ containing $M\times M$  implicit-contact subgraphs are rejected at the $M$-rule level.

\begin{figure}[htbp]
\includegraphics[width=2.75in]{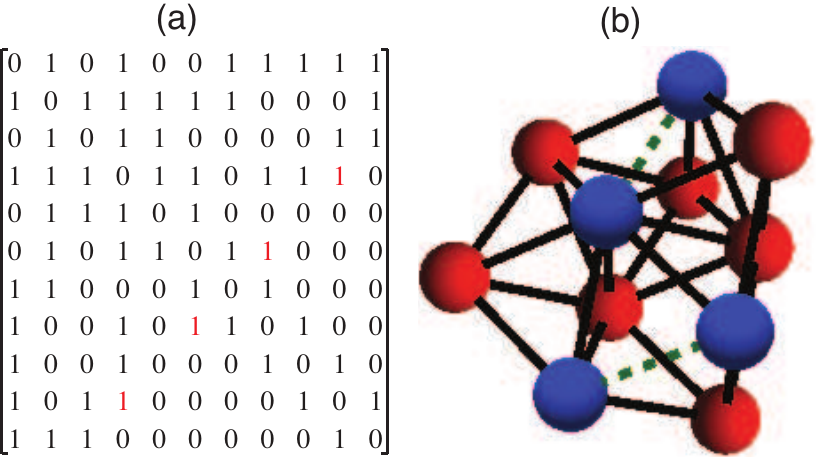}
\caption{(Color online) Panel (a): An implicit contact adjacency matrix with $M=11$, 27 explicit contacts (black 1's) and two implicit contacts arising from the solution of Equation \ref{eq:cartesianset} (red [lighter shaded] 1's).  Panel (b): Visualizing the structure reveals a $N_c = 29$ packing; implicit contacts are shown as dashed green lines.}
\label{fig:implic29ill}
\end{figure}

We find 29 nonisomorphic $M = 10$, $M_c = 24$ implicit-contact graphs.
All reduce to one of the three $M = 10$, $M_c = 25$ minimal energy packings \cite{arkus09} when we solve for $\{\vec{r}\}$. 
Note that we overcounted $\mathcal{M}(10,24)$ in Ref.\ \cite{hoy10} by including 20 implicit-contact macrostates.  
Similarly, we find 13 $M = 11$, $M_c = 27$ graphs, not containing isomorphic subgraphs of the abovementioned set of 10/24 graphs, that imply structures with 28 or more contacts \cite{onlinepackings}.
Fig.\ \ref{fig:implic29ill} illustrates one such structure: panel (a) shows an implicit contact adjacency matrix with $N = 11$ and $N_c = 27$; contacts present in $\bar{A}$ are shown in black.
The corresponding packing (panel (b)) possesses 29 contacts (as determined by solving for $\{\vec{r}\}$) and is identical to the $N = 11, N_c = 29$ packing shown in Fig.\ \ref{fig:mostleast}(a). 
The implicit contacts are shown in red in panel (a).
In panel (b), the implicit contacts are shown as dashed lines, and spheres possessing implicit contacts (particles 4, 6, 8, and 10) are shaded blue (dark).

\end{appendix}


\begin{thebibliography}{78}
\expandafter\ifx\csname natexlab\endcsname\relax\def\natexlab#1{#1}\fi
\expandafter\ifx\csname bibnamefont\endcsname\relax
  \def\bibnamefont#1{#1}\fi
\expandafter\ifx\csname bibfnamefont\endcsname\relax
  \def\bibfnamefont#1{#1}\fi
\expandafter\ifx\csname citenamefont\endcsname\relax
  \def\citenamefont#1{#1}\fi
\expandafter\ifx\csname url\endcsname\relax
  \def\url#1{\texttt{#1}}\fi
\expandafter\ifx\csname urlprefix\endcsname\relax\def\urlprefix{URL }\fi
\providecommand{\bibinfo}[2]{#2}
\providecommand{\eprint}[2][]{\url{#2}}

\bibitem[{\citenamefont{Hales}(1998)}]{hales98}
\bibinfo{author}{\bibfnamefont{T.~C.} \bibnamefont{Hales}},
  \bibinfo{journal}{Ann. Math.} \textbf{\bibinfo{volume}{162}},
  \bibinfo{pages}{1065} (\bibinfo{year}{1998}).
  
\bibitem[{\citenamefont{Berryman}(1983)}]{berryman83}
\bibinfo{author}{\bibfnamefont{J.~G.} \bibnamefont{Berryman}},
  \bibinfo{journal}{Phys. Rev. A} \textbf{\bibinfo{volume}{27}},
  \bibinfo{pages}{1053} (\bibinfo{year}{1983}).  

\bibitem[{\citenamefont{Rintoul and Torquato}(1996)}]{rintoul96}
\bibinfo{author}{\bibfnamefont{M.~D.} \bibnamefont{Rintoul}} \bibnamefont{and}
  \bibinfo{author}{\bibfnamefont{S.}~\bibnamefont{Torquato}},
  \bibinfo{journal}{J. Chem. Phys.} \textbf{\bibinfo{volume}{105}},
  \bibinfo{pages}{9258} (\bibinfo{year}{1996}).

\bibitem[{\citenamefont{Torquato et~al.}(2000)\citenamefont{Torquato, Truskett,
  and Debenedetti}}]{torquato00}
\bibinfo{author}{\bibfnamefont{S.}~\bibnamefont{Torquato}},
  \bibinfo{author}{\bibfnamefont{T.~M.} \bibnamefont{Truskett}},
  \bibnamefont{and} \bibinfo{author}{\bibfnamefont{P.~G.}
  \bibnamefont{Debenedetti}}, \bibinfo{journal}{Phys. Rev. Lett.}
  \textbf{\bibinfo{volume}{84}}, \bibinfo{pages}{2064} (\bibinfo{year}{2000}).

\bibitem[{\citenamefont{O'Hern et~al.}(2003)\citenamefont{O'Hern, Silbert, Liu,
  and Nagel}}]{ohern03}
\bibinfo{author}{\bibfnamefont{C.~S.} \bibnamefont{O'Hern}},
  \bibinfo{author}{\bibfnamefont{L.~E.} \bibnamefont{Silbert}},
  \bibinfo{author}{\bibfnamefont{A.~J.} \bibnamefont{Liu}}, \bibnamefont{and}
  \bibinfo{author}{\bibfnamefont{S.~R.} \bibnamefont{Nagel}},
  \bibinfo{journal}{Phys. Rev. E} \textbf{\bibinfo{volume}{68}},
  \bibinfo{pages}{011306} (\bibinfo{year}{2003}).

\bibitem[{\citenamefont{Torquato and Stillinger}(2010)}]{torquato10}
\bibinfo{author}{\bibfnamefont{S.}~\bibnamefont{Torquato}} \bibnamefont{and}
  \bibinfo{author}{\bibfnamefont{F.~H.} \bibnamefont{Stillinger}},
  \bibinfo{journal}{Rev. Mod. Phys.} \textbf{\bibinfo{volume}{82}},
  \bibinfo{pages}{2633} (\bibinfo{year}{2010}).

\bibitem[{\citenamefont{Foffi et~al.}(2000)\citenamefont{Foffi, Zaccarelli,
  Sciortino, and Tartaglia}}]{foffi00}
\bibinfo{author}{\bibfnamefont{G.}~\bibnamefont{Foffi}},
  \bibinfo{author}{\bibfnamefont{E.}~\bibnamefont{Zaccarelli}},
  \bibinfo{author}{\bibfnamefont{F.}~\bibnamefont{Sciortino}},
  \bibnamefont{and}
  \bibinfo{author}{\bibfnamefont{P.}~\bibnamefont{Tartaglia}},
  \bibinfo{journal}{J. Stat. Phys.} \textbf{\bibinfo{volume}{100}},
  \bibinfo{pages}{363} (\bibinfo{year}{2000});
\bibinfo{author}{\bibfnamefont{G.}~\bibnamefont{Foffi}}~\textit{et.al},
  \bibinfo{journal}{Phys. Rev. E} \textbf{\bibinfo{volume}{65}},
  \bibinfo{pages}{031407} (\bibinfo{year}{2002}).  

\bibitem[{\citenamefont{Hopkins et~al.}(2011)\citenamefont{Hopkins, Stillinger,
  and Torquato}}]{hopkins11}
\bibinfo{author}{\bibfnamefont{A.~B.} \bibnamefont{Hopkins}},
  \bibinfo{author}{\bibfnamefont{F.~H.} \bibnamefont{Stillinger}},
  \bibnamefont{and} \bibinfo{author}{\bibfnamefont{S.}~\bibnamefont{Torquato}},
  \bibinfo{journal}{Phys. Rev. E} \textbf{\bibinfo{volume}{83}}
  (\bibinfo{year}{2011}).

\bibitem[{\citenamefont{Barlow}(1883)}]{barlow1883}
\bibinfo{author}{\bibfnamefont{W.}~\bibnamefont{Barlow}},
  \bibinfo{journal}{Nature} \textbf{\bibinfo{volume}{29}}, \bibinfo{pages}{186}
  (\bibinfo{year}{1883}).

\bibitem[{bar()}]{barlowfoot}
\bibinfo{note}{In the remainder of this paper, we use the term ``Barlow'' in place of
  ``random hexagonal close packed'' since the latter term can exclude
  packings with FCC or HCP order. See Section \ref{subsec:structanaly}.}

\bibitem[{\citenamefont{Meng et~al.}(2010)\citenamefont{Meng, Arkus, Brenner,
  and Manoharan}}]{meng10}
\bibinfo{author}{\bibfnamefont{G.}~\bibnamefont{Meng}},
  \bibinfo{author}{\bibfnamefont{N.}~\bibnamefont{Arkus}},
  \bibinfo{author}{\bibfnamefont{M.~P.} \bibnamefont{Brenner}},
  \bibnamefont{and} \bibinfo{author}{\bibfnamefont{V.~N.}
  \bibnamefont{Manoharan}}, \bibinfo{journal}{Science}
  \textbf{\bibinfo{volume}{327}}, \bibinfo{pages}{560} (\bibinfo{year}{2010}).

\bibitem[{\citenamefont{O'Malley and Snook}(2003)}]{omalley03}
\bibinfo{author}{\bibfnamefont{B.}~\bibnamefont{O'Malley}} \bibnamefont{and}
  \bibinfo{author}{\bibfnamefont{I.}~\bibnamefont{Snook}},
  \bibinfo{journal}{Phys. Rev. Lett.} \textbf{\bibinfo{volume}{90}},
  \bibinfo{pages}{085702} (\bibinfo{year}{2003}).

\bibitem[{\citenamefont{Zaccarelli et~al.}(2009)\citenamefont{Zaccarelli,
  Valeriani, Sanz, Poon, Cates, and Pusey}}]{zaccarrelli09}
\bibinfo{author}{\bibfnamefont{E.}~\bibnamefont{Zaccarelli}},
  \bibinfo{author}{\bibfnamefont{C.}~\bibnamefont{Valeriani}},
  \bibinfo{author}{\bibfnamefont{E.}~\bibnamefont{Sanz}},
  \bibinfo{author}{\bibfnamefont{W.~C.~K.}~\bibnamefont{Poon}},
  \bibinfo{author}{\bibfnamefont{M.~E.}~\bibnamefont{Cates}}, \bibnamefont{and}
  \bibinfo{author}{\bibfnamefont{P.~N.} \bibnamefont{Pusey}},
  \bibinfo{journal}{Phys. Rev. Lett.} \textbf{\bibinfo{volume}{103}},
  \bibinfo{pages}{135704} (\bibinfo{year}{2009}).

\bibitem[{\citenamefont{Karayiannis et~al.}(2011)\citenamefont{Karayiannis,
  Malshe, {de Pablo}, and Laso}}]{karayiannis11}
\bibinfo{author}{\bibfnamefont{N.~C.}~\bibnamefont{Karayiannis}},
  \bibinfo{author}{\bibfnamefont{R.}~\bibnamefont{Malshe}},
  \bibinfo{author}{\bibfnamefont{J.~J.}~\bibnamefont{{de Pablo}}},
  \bibnamefont{and} \bibinfo{author}{\bibfnamefont{M.}~\bibnamefont{Laso}},
  \bibinfo{journal}{Phys. Rev. E} \textbf{\bibinfo{volume}{83}},
  \bibinfo{pages}{061605} (\bibinfo{year}{2011});
\bibinfo{author}{\bibfnamefont{N.~C.}~\bibnamefont{Karayiannis}},
  \bibinfo{author}{\bibfnamefont{R.}~\bibnamefont{Malshe}},
  \bibinfo{author}{\bibfnamefont{M.}~\bibnamefont{Kr{\"o}ger}},
  \bibinfo{author}{\bibfnamefont{J.~J.}~\bibnamefont{{de Pablo}}},
  \bibnamefont{and} \bibinfo{author}{\bibfnamefont{M.}~\bibnamefont{Laso}},
  \bibinfo{journal}{Soft Matter} \textbf{\bibinfo{volume}{8}},
  \bibinfo{pages}{844} (\bibinfo{year}{2012}).

\bibitem[{\citenamefont{Schillling, Schope, Oettel, Optelal, and Snook}(2010)}]{schilling10}
\bibinfo{author}{\bibfnamefont{T.} \bibnamefont{Schilling}},
\bibinfo{author}{\bibfnamefont{H.~J.} \bibnamefont{Sch{\"o}pe}},
\bibinfo{author}{\bibfnamefont{M.} \bibnamefont{Oettell}},
\bibinfo{author}{\bibfnamefont{G.} \bibnamefont{Optelal}}, \bibnamefont{and}
\bibinfo{author}{\bibfnamefont{I.} \bibnamefont{Snook}},
 \bibinfo{journal}{Phys. Rev. Lett.} \textbf{\bibinfo{volume}{105}}, \bibinfo{pages}{025701} (\bibinfo{year}{2010}).

\bibitem[{\citenamefont{Taffs,Malins,Williams, and Royall}(2010)}]{taffs10}
\bibinfo{author}{\bibfnamefont{J.} \bibnamefont{Taffs}}, 
\bibinfo{author}{\bibfnamefont{T.} \bibnamefont{Ohtsuka}}
\bibinfo{author}{\bibfnamefont{S.~R.} \bibnamefont{Williams}}, \bibnamefont{and}
\bibinfo{author}{\bibfnamefont{C.~P.} \bibnamefont{Royall}},
 \bibinfo{journal}{J. Chem. Phys.} \textbf{\bibinfo{volume}{133}},
  \bibinfo{pages}{244901} (\bibinfo{year}{2010}).

\bibitem[{\citenamefont{Miller et~al.}(2004)\citenamefont{Miller and Frenkel
  and Pisanski}}]{miller04}
\bibinfo{author}{\bibfnamefont{M.~A.} \bibnamefont{Miller}} \bibnamefont{and}
\bibinfo{author}{\bibfnamefont{D.}~\bibnamefont{Frenkel}},
  \bibinfo{journal}{Phys. Rev. Lett.} \textbf{\bibinfo{volume}{90}},
  \bibinfo{pages}{135702} (\bibinfo{year}{2003});  
\bibinfo{author}{\bibfnamefont{M.~A.} \bibnamefont{Miller}} \bibnamefont{and}
\bibinfo{author}{\bibfnamefont{D.}~\bibnamefont{Frenkel}},
  \bibinfo{journal}{J. Phys. Cond. Matt.} \textbf{\bibinfo{volume}{16}},
  \bibinfo{pages}{S4901} (\bibinfo{year}{2004});
\bibinfo{author}{\bibfnamefont{P.}~\bibnamefont{Charbonneau}} \bibnamefont{and}
\bibinfo{author}{\bibfnamefont{D.}~\bibnamefont{Frenkel}},
  \bibinfo{journal}{J. Chem. Phys.} \textbf{\bibinfo{volume}{126}},
  \bibinfo{pages}{196101} (\bibinfo{year}{2007}).
  
\bibitem[{\citenamefont{Auer and Frenkel}(2001)}]{auer01}
\bibinfo{author}{\bibfnamefont{S.}~\bibnamefont{Auer}} \bibnamefont{and}
  \bibinfo{author}{\bibfnamefont{D.}~\bibnamefont{Frenkel}},
  \bibinfo{journal}{Nature} \textbf{\bibinfo{volume}{409}},
  \bibinfo{pages}{1020} (\bibinfo{year}{2001});
\bibinfo{author}{\bibfnamefont{S.}~\bibnamefont{Auer}} \bibnamefont{and}
  \bibinfo{author}{\bibfnamefont{D.}~\bibnamefont{Frenkel}},
  \bibinfo{journal}{Ann. Rev. Phys. Chem.} \textbf{\bibinfo{volume}{55}},
  \bibinfo{pages}{333} (\bibinfo{year}{2004}).

\bibitem[{\citenamefont{Crocker}(2010)}]{crocker10}
\bibinfo{author}{\bibfnamefont{J.~C.} \bibnamefont{Crocker}},
  \bibinfo{journal}{Science} \textbf{\bibinfo{volume}{327}},
  \bibinfo{pages}{535} (\bibinfo{year}{2010}).

\bibitem[{\citenamefont{Buzzacaro et~al.}(2007)\citenamefont{Buzzacaro,
  Rusconi, and Piazza}}]{buzzacaro07}
\bibinfo{author}{\bibfnamefont{S.}~\bibnamefont{Buzzacaro}},
  \bibinfo{author}{\bibfnamefont{R.}~\bibnamefont{Rusconi}}, \bibnamefont{and}
  \bibinfo{author}{\bibfnamefont{R.}~\bibnamefont{Piazza}},
  \bibinfo{journal}{Phys. Rev. Lett.} \textbf{\bibinfo{volume}{99}},
  \bibinfo{pages}{098301} (\bibinfo{year}{2007}).  

\bibitem[{\citenamefont{Hoare and McInnes}(1976)}]{hoare76}
\bibinfo{author}{\bibfnamefont{M.~R.} \bibnamefont{Hoare}} \bibnamefont{and}
  \bibinfo{author}{\bibfnamefont{J.}~\bibnamefont{McInnes}},
  \bibinfo{journal}{Faraday Discuss. Chem. Soc.} \textbf{\bibinfo{volume}{61}},
  \bibinfo{pages}{12} (\bibinfo{year}{1976}).

\bibitem[{\citenamefont{Arkus et~al.}(2009)\citenamefont{Arkus, Manoharan, and
  Brenner}}]{arkus09}
\bibinfo{author}{\bibfnamefont{N.}~\bibnamefont{Arkus}},
  \bibinfo{author}{\bibfnamefont{V.~N.} \bibnamefont{Manoharan}},
  \bibnamefont{and} \bibinfo{author}{\bibfnamefont{M.~P.}
  \bibnamefont{Brenner}}, \bibinfo{journal}{Phys. Rev. Lett.}
  \textbf{\bibinfo{volume}{103}}, \bibinfo{pages}{118303}
  (\bibinfo{year}{2009}).

\bibitem[{\citenamefont{Bernal}(1960)}]{bernal60}
\bibinfo{author}{\bibfnamefont{J.~D.} \bibnamefont{Bernal}},
  \bibinfo{journal}{Nature} \textbf{\bibinfo{volume}{188}},
  \bibinfo{pages}{910} (\bibinfo{year}{1960}).

\bibitem[{\citenamefont{Conway and Sloane}(1998)}]{conway98}
\bibinfo{author}{\bibfnamefont{J.~H.} \bibnamefont{Conway}} \bibnamefont{and}
  \bibinfo{author}{\bibfnamefont{N.~J.~H.} \bibnamefont{Sloane}},
  \emph{\bibinfo{title}{Sphere Packings, Lattices, and Groups}}
  (\bibinfo{publisher}{Springer}, \bibinfo{year}{1998}).

\bibitem[{\citenamefont{Arkus et~al.}(2011)\citenamefont{Arkus, Manoharan, and
  Brenner}}]{arkus11}
\bibinfo{author}{\bibfnamefont{N.}~\bibnamefont{Arkus}},
  \bibinfo{author}{\bibfnamefont{V.~N.} \bibnamefont{Manoharan}},
  \bibnamefont{and} \bibinfo{author}{\bibfnamefont{M.~P.}
  \bibnamefont{Brenner}}, \bibinfo{journal}{Siam J. Discrete Math.}
  \textbf{\bibinfo{volume}{25}}, \bibinfo{pages}{1860} (\bibinfo{year}{2011}).

\bibitem[{\citenamefont{Wales}(2004)}]{wales04}
\bibinfo{author}{\bibfnamefont{D.~J.} \bibnamefont{Wales}},
  \emph{\bibinfo{title}{Energy Landscapes: Applications to Clusters,
  Biomolecules and Glasses}} (\bibinfo{publisher}{Cambridge Molecular Science},
  \bibinfo{year}{2004}).

\bibitem[{\citenamefont{Stell}(1991)}]{stell91}
\bibinfo{author}{\bibfnamefont{G.}~\bibnamefont{Stell}}, \bibinfo{journal}{J.
  Stat. Phys.} \textbf{\bibinfo{volume}{63}}, \bibinfo{pages}{1203}
  (\bibinfo{year}{1991}).

\bibitem[{\citenamefont{{Hansen-Goos} and Wettlaufer}(2011)}]{hhgw11}
\bibinfo{author}{\bibfnamefont{H.}~\bibnamefont{{Hansen-Goos}}}
  \bibnamefont{and} \bibinfo{author}{\bibfnamefont{J.~S.}
  \bibnamefont{Wettlaufer}}, \bibinfo{journal}{J. Chem. Phys.}
  \textbf{\bibinfo{volume}{134}}, \bibinfo{pages}{014506}
  (\bibinfo{year}{2011}).

\bibitem[{\citenamefont{Mossa and Tarjus}(2003)}]{mossa03}
\bibinfo{author}{\bibfnamefont{S.} \bibnamefont{Mossa}} \bibnamefont{and} 
\bibinfo{author}{\bibfnamefont{G.} \bibnamefont{Tarjus}},
  \bibinfo{journal}{J. Chem. Phys.} \textbf{\bibinfo{volume}{22}},
  \bibinfo{pages}{8070} (\bibinfo{year}{2003});
\bibinfo{author}{\bibfnamefont{R.~L.} \bibnamefont{Davidchack}} \bibnamefont{and} 
\bibinfo{author}{\bibfnamefont{B.~B.} \bibnamefont{Laird}},
  \bibinfo{journal}{Phys. Rev. Lett.} \textbf{\bibinfo{volume}{94}},
  \bibinfo{pages}{086102} (\bibinfo{year}{2005})

\bibitem[{\citenamefont{Richards}(1977)}]{richards77}
\bibinfo{author}{\bibfnamefont{F.~M.} \bibnamefont{Richards}},
  \bibinfo{journal}{Ann. Rev. Biophys. Bioeng.} \textbf{\bibinfo{volume}{6}},
  \bibinfo{pages}{151} (\bibinfo{year}{1977}).

\bibitem[{\citenamefont{ten Wolde and Frenkel}(1997))}]{tenwolde97} 
\bibinfo{author}{\bibfnamefont{P.~R.} \bibnamefont{ten Wolde}} \bibnamefont{and}
  \bibinfo{author}{\bibfnamefont{D.} \bibnamefont{Frenkel}},
   \bibinfo{journal}{Science} \textbf{\bibinfo{volume}{277}},
  \bibinfo{pages}{1975} (\bibinfo{year}{1997}).

\bibitem[{\citenamefont{Miller et~al.}(1997)\citenamefont{Miller, Teng,
  Thurston, and Vavasis}}]{miller97}
\bibinfo{author}{\bibfnamefont{G.~L.} \bibnamefont{Miller}},
  \bibinfo{author}{\bibfnamefont{S.}~\bibnamefont{Teng}},
  \bibinfo{author}{\bibfnamefont{W.}~\bibnamefont{Thurston}}, \bibnamefont{and}
  \bibinfo{author}{\bibfnamefont{S.~A.} \bibnamefont{Vavasis}},
  \bibinfo{journal}{Journal of the ACM} \textbf{\bibinfo{volume}{44}},
  \bibinfo{pages}{1} (\bibinfo{year}{1997}).

\bibitem[{\citenamefont{Johnson}(1962)}]{johnson62}
\bibinfo{author}{\bibfnamefont{S.}~\bibnamefont{Johnson}},
  \bibinfo{journal}{IRE Transactions on Information Theory}
  \textbf{\bibinfo{volume}{8}}, \bibinfo{pages}{203} (\bibinfo{year}{1962}).

\bibitem[{\citenamefont{Erdos}(1946)}]{erdos46}
\bibinfo{author}{\bibfnamefont{P.}~\bibnamefont{Erd\H{o}s}}, \bibinfo{journal}{Am.
  Math. Monthly} \textbf{\bibinfo{volume}{53}}, \bibinfo{pages}{248}
  (\bibinfo{year}{1946}).

\bibitem[{\citenamefont{Horvat et~al.}(2011)\citenamefont{Horvat, Kratochvil,
  and Pisanski}}]{horvat11}
\bibinfo{author}{\bibfnamefont{B.}~\bibnamefont{Horvat}},
  \bibinfo{author}{\bibfnamefont{J.}~\bibnamefont{Kratochvil}},
  \bibnamefont{and} \bibinfo{author}{\bibfnamefont{T.}~\bibnamefont{Pisanski}},
  \bibinfo{journal}{Lect. Notes Computer Sci.} \textbf{\bibinfo{volume}{6460}},
  \bibinfo{pages}{274} (\bibinfo{year}{2011}).

\bibitem[{foo({\natexlab{a}})}]{footMC}
\bibinfo{note}{For example, if one places spheres in a box of volume $V$ and
  uses a typical ``displacement move'' of size $\eta$, the number of particle
  moves required to determine $\mathcal{M}(N, N_c)$ scales as
  $N\exp{(3NV/\eta^{3})}$ - and of course $\eta$ must be miniscule to obtain a
  robust solution.}

\bibitem[{\citenamefont{Hoy and O'Hern}(2010)}]{hoy10}
\bibinfo{author}{\bibfnamefont{R.~S.} \bibnamefont{Hoy}} \bibnamefont{and}
  \bibinfo{author}{\bibfnamefont{C.~S.} \bibnamefont{O'Hern}},
  \bibinfo{journal}{Phys. Rev. Lett.} \textbf{\bibinfo{volume}{105}},
  \bibinfo{pages}{068001} (\bibinfo{year}{2010}).

\bibitem[{foo({\natexlab{b}})}]{footfloppy}
\bibinfo{note}{We consider both mechanically stable and floppy packings. Stable
  packings correspond to zero-dimensional points in configuration space. Floppy
  packings occupy finite volumes in configuration space, but we have verified
  that these are disconnected and correspond to distinguishable inherent
  structures \cite{stillinger95}.}
 
  
\bibitem[{\citenamefont{Stillinger}(1995)}]{stillinger95}
\bibinfo{author}{\bibfnamefont{F.~H.} \bibnamefont{Stillinger}},
  \bibinfo{journal}{Science} \textbf{\bibinfo{volume}{267}},
  \bibinfo{pages}{1935} (\bibinfo{year}{1995}).

\bibitem[{\citenamefont{Yuste and Santos}(1993)}]{yuste93}
\bibinfo{author}{\bibfnamefont{S.~B.} \bibnamefont{Yuste}} \bibnamefont{and}
  \bibinfo{author}{\bibfnamefont{A.}~\bibnamefont{Santos}},
  \bibinfo{journal}{Phys. Rev. E} \textbf{\bibinfo{volume}{48}},
  \bibinfo{pages}{4599} (\bibinfo{year}{1993}).  

\bibitem[{\citenamefont{Baxter}(1968)}]{baxter68}
\bibinfo{author}{\bibfnamefont{R.~J.} \bibnamefont{Baxter}},
  \bibinfo{journal}{J. Chem. Phys.} \textbf{\bibinfo{volume}{49}},
  \bibinfo{pages}{2770} (\bibinfo{year}{1968}).


\bibitem[{foo({\natexlab{c}})}]{footchiral}
\bibinfo{note}{Chiral enantiomer pairs related by a mirror symmetry have the
  same $\{r_{ij}^2\}$ and isomorphic $\bar{A}$. Following the convention of
  Refs.\ \cite{arkus09,meng10,arkus11}, we count enantiomer pairs as single
  macrostates. Note that our Euclidean structure solver (Section
  \ref{subsec:euclidean}) finds left- and right-handed enantiomers with equal
  probability.}

\bibitem[{\citenamefont{Wales}(2010)}]{wales10}
\bibinfo{author}{\bibfnamefont{D.~J.} \bibnamefont{Wales}},
  \bibinfo{journal}{ChemPhysChem} \textbf{\bibinfo{volume}{11}},
  \bibinfo{pages}{2491} (\bibinfo{year}{2010});
\bibinfo{author}{\bibfnamefont{F.}~\bibnamefont{Calvo}},
\bibinfo{author}{\bibfnamefont{J.~P.~K.} \bibnamefont{Doye}}, \bibnamefont{and}
\bibinfo{author}{\bibfnamefont{D.~J.} \bibnamefont{Wales}},
  \bibinfo{journal}{Nanoscale} \textbf{\bibinfo{volume}{4}},
  \bibinfo{pages}{1085}  (\bibinfo{year}{2012}).
  


\bibitem[{\citenamefont{Jacobs and Thorpe}(1995)}]{jacobs95}
\bibinfo{author}{\bibfnamefont{D.~J.} \bibnamefont{Jacobs}} \bibnamefont{and}
  \bibinfo{author}{\bibfnamefont{M.~F.} \bibnamefont{Thorpe}},
  \bibinfo{journal}{Phys. Rev. Lett.} \textbf{\bibinfo{volume}{75}},
  \bibinfo{pages}{4051} (\bibinfo{year}{1995}).

\bibitem[{foo({\natexlab{d}})}]{footbinary}
\bibinfo{note}{The $i^{th}$ digit of $\mathcal{B}$ is the $i^{th}$ element of
  $\bar{A}$, where elements above the diagonal are ordered left-to-right and
  top-to bottom. The enumeration over all $\mathcal{B}$ can be efficiently
  executed using the C++ Standard Template Library's $next\_permutation()$
  function, which our code employs.}

\bibitem[{\citenamefont{{Biedl \textrm{et.\ al.}}}(2001)}]{biedl01}
\bibinfo{author}{\bibfnamefont{T.}~\bibnamefont{{Biedl \textrm{et.\ al.}}}},
  \bibinfo{journal}{Discrete Comput. Geom.} \textbf{\bibinfo{volume}{26}},
  \bibinfo{pages}{269} (\bibinfo{year}{2001}).

\bibitem[{foo({\natexlab{e}})}]{footAk}
\bibinfo{note}{$A_k$ is the number of particle index permutations mapping an
  adjacency matrix onto itself while preserving edge-vertex connectivity. We
  evaluate $\mathcal{A}_k$ using Mathematica$^{TM}$.}

\bibitem[{foo({\natexlab{f}})}]{footdisting}
\bibinfo{note}{Arguments based on the foundations of statistical mechanics
  \cite{swendsen02} suggest that few-body classical systems in which one is
  \textit{able} to distinguish the particles should indeed be treated as
  composed of distinguishable particles and that doing so produces no
  anomalies. For example, real systems \cite{buzzacaro07} are inevitably
  polydisperse, and experiments employing techniques such as optical microscopy
  \cite{meng10} can track the motion of individual particles.}

\bibitem[{\citenamefont{Swendsen}(2002)}]{swendsen02}
\bibinfo{author}{\bibfnamefont{R.~H.} \bibnamefont{Swendsen}},
  \bibinfo{journal}{J. Stat. Phys.} \textbf{\bibinfo{volume}{107}},
  \bibinfo{pages}{1143} (\bibinfo{year}{2002}).
  
\bibitem[{foo({\natexlab{g}})}]{footribi}
\bibinfo{note}{The triangular bipyramid rule \cite{arkus11} is effective in
  determining unknown distances in $N$-particle, $N_c$-contact packings that
  are \textit{iterative} (i.e.\ for $N > M$ and $M_c < N_c-3$, contain a subset
  of the $\mathcal{M}(M,M_c)$ stable packings). However, it fails for
  noniterative ``new seed'' packings \cite{arkus09,arkus11} as well as packings
  including linear trimers. The fractions ot both noniterative and linear-trimer-containing 
  nuclei increase with increasing $N$.}

\bibitem[{kur()}]{kuratowski}
\bibinfo{note}{{h}ttp://en.wikipedia.org/wiki/Planar$\_$graph}.

\bibitem[{hli()}]{hlinenythesis}
\bibinfo{note}{P. Hlin{\u{e}}n{\'y}, Ph.\ D.\ thesis, Charles University in
  Prague. 2000.}

\bibitem[{\citenamefont{Harborth et~al.}(2002)\citenamefont{Harborth,
  Szab{\'o}, and {Ujv{\'a}ry-Menyh{\'a}rt}}}]{harborth02}
\bibinfo{author}{\bibfnamefont{H.}~\bibnamefont{Harborth}},
  \bibinfo{author}{\bibfnamefont{L.}~\bibnamefont{Szab{\'o}}},
  \bibnamefont{and}
  \bibinfo{author}{\bibfnamefont{Z.}~\bibnamefont{{Ujv{\'a}ry-Menyh{\'a}rt}}},
  \bibinfo{journal}{Arch. Math} \textbf{\bibinfo{volume}{78}},
  \bibinfo{pages}{81} (\bibinfo{year}{2002}).

\bibitem[{foo({\natexlab{h}})}]{footoctexcep}
\bibinfo{note}{An exception is the $M=6, M_c = 12$ octahedron shown in Fig.\
  \ref{fig:macrostateexamp}(a), but this structure is found in FCC and HCP
  lattices as well.}

\bibitem[{foo({\natexlab{i}})}]{footWein}
\bibinfo{note}{To see this, consider a set of ``local'' rules sufficient to
  reject all invalid $N$-sphere packings. One can map any two valid $N$-sphere
  packings to a packing of $2N$ particles which violates none of the local
  rules yet is invalid because it implies two particles $\{i,j\}$ are
  conincident ($\vec{r}_i = \vec{r}_j$). This argument is based on private
  communication of an unpublished result of Shmuel Weinberger, Dept.\ of
  Mathematics, University of Chicago.}

\bibitem[{\citenamefont{Dahlquist and Bjorck}(1974)}]{dahlquist74}
\bibinfo{author}{\bibfnamefont{G.}~\bibnamefont{Dahlquist}} \bibnamefont{and}
  \bibinfo{author}{\bibfnamefont{A.}~\bibnamefont{Bjorck}},
  \emph{\bibinfo{title}{Numerical Methods}}
  (\bibinfo{publisher}{Prentice-Hall}, \bibinfo{year}{1974}).

\bibitem[{sha()}]{shattuck}
\bibinfo{note}{T.\ W.\ Shattuck, ABC Rotational ConstantCalculator,
  http://www.colby.edu/chemistry/PChem/scripts/ ABC.html.}

\bibitem[{\citenamefont{{ten Wolde} et~al.}(1996)\citenamefont{{ten Wolde},
  {Ruiz-Montero}, and Frenkel}}]{tenwolde96}
\bibinfo{author}{\bibfnamefont{P.~R.} \bibnamefont{{ten Wolde}}},
  \bibinfo{author}{\bibfnamefont{M.~J.} \bibnamefont{{Ruiz-Montero}}},
  \bibnamefont{and} \bibinfo{author}{\bibfnamefont{D.}~\bibnamefont{Frenkel}},
  \bibinfo{journal}{J. Chem. Phys.} \textbf{\bibinfo{volume}{104}},
  \bibinfo{pages}{9932} (\bibinfo{year}{1996}).

\bibitem[{\citenamefont{Stillinger and Weber}(1982)}]{stillinger82}
\bibinfo{author}{\bibfnamefont{F.~H.} \bibnamefont{Stillinger}}
  \bibnamefont{and} \bibinfo{author}{\bibfnamefont{T.~A.} \bibnamefont{Weber}},
  \bibinfo{journal}{Phys. Rev. A} \textbf{\bibinfo{volume}{25}},
  \bibinfo{pages}{978} (\bibinfo{year}{1982}).


\bibitem[{foo({\natexlab{j}})}]{footgao}
\bibinfo{note}{It has been demonstrated for bidisperse 2D disk packings; G.-J.
  Gao, J.\ Blawzdziewicz, and C.\ S.\ O'Hern. Phys. Rev. E \textbf{80}, 061303
  (2009).}

\bibitem[{\citenamefont{Phillips}(1979)}]{phillips79}
\bibinfo{author}{\bibfnamefont{J.~C.} \bibnamefont{Phillips}},
  \bibinfo{journal}{J. Non-Cryst. Solids} \textbf{\bibinfo{volume}{34}},
  \bibinfo{pages}{153} (\bibinfo{year}{1979});
\bibinfo{author}{\bibfnamefont{M.~F.} \bibnamefont{Thorpe}},
  \bibinfo{journal}{J. Non-Cryst. Solids} \textbf{\bibinfo{volume}{57}},
  \bibinfo{pages}{355} (\bibinfo{year}{1983}).

\bibitem[{\citenamefont{Royall, Williams, Ohtsuka and Tanaka}(2008)}]{royall08}
\bibinfo{author}{\bibfnamefont{C.~P.} \bibnamefont{Royall}},
\bibinfo{author}{\bibfnamefont{S.~R.} \bibnamefont{Williams}},
\bibinfo{author}{\bibfnamefont{T.} \bibnamefont{Ohtsuka}}, \bibnamefont{and}
\bibinfo{author}{\bibfnamefont{H.} \bibnamefont{Tanaka}}, 
 \bibinfo{journal}{Nature Materials} \textbf{\bibinfo{volume}{7}},
  \bibinfo{pages}{556} (\bibinfo{year}{2008}).



\bibitem[{foo({\natexlab{k}})}]{footnobridge}
\bibinfo{note}{Since we focus on nuclei satisfying the minimal stability
  criteria \textbf{(i-ii)}, results for $N = 11,\ N_c = 27$ throughout this
  section exclude the ``bridge'' packings described in Appendix
  \ref{sec:bridge}.}

\bibitem[{foo({\natexlab{k}})}]{onlinepackings}
\bibinfo{note}{Adjacency matrices $\bar{A}$ and particle coordinates $\vec{r}$ for all nonisomorphic packings reported here, as well as adjacency matrices for the implicit-contact graphs discussed in Appendix \ref{sec:implic}, are available for download at http://jamming.research.yale.edu/data/\\ stickyhardspherepackings.tar.gz.}

\bibitem[{foo({\natexlab{m}})}]{footclassicalN}
\bibinfo{note}{At sufficiently large $N \sim N_{classical}$, the minimal energy
  clusters of sticky hard spheres will be close-packed and defect-free, and
  classical nucleation theory should perform better. However, $N_{classical}$
  is unknown. Ref.\ \cite{arkus11} shows that MEPs for several $N\geq14$ do not
  possess FCC, HCP, or Barlow order but instead are stack-faulted. Ref.\
  \cite{hopkins11} showed that the densest spherical packings retain non-Barlow
  order for $N \lesssim 500$.}

\bibitem[{foo({\natexlab{n}})}]{footoct}
\bibinfo{note}{For $N=6$, $r^{min}_{2nd} = 2\sqrt{2}/3 > \sqrt{2}$; this
  distance is found in the octahedron, which is is a subsection of both the FCC
  and HCP lattices.}

\bibitem[{foo({\natexlab{l}})}]{foottech}
\bibinfo{note}{Note that stack-faulted structures (e.g.\ the capped trigonal
  bipyramid structure in Fig.\ \ref{fig:macrostateexamp}(b)) to not in general
  have either $C_1$ or Barlow ordering and that some structures incompatible
  with close-packed ordering, i.e.\ 5-fold symmetric structures (see Section
  \ref{subsec:hostilemotifs}) have higher symmetry than $C_1$.}


\bibitem[{\citenamefont{Cochran and Chiew}(2006)}]{cochran06}
\bibinfo{author}{\bibfnamefont{T.~W.} \bibnamefont{Cochran}} \bibnamefont{and}
  \bibinfo{author}{\bibfnamefont{Y.~C.} \bibnamefont{Chiew}},
  \bibinfo{journal}{J. Chem. Phys.} \textbf{\bibinfo{volume}{124}},
  \bibinfo{pages}{224901} (\bibinfo{year}{2006}).

\bibitem[{foo({\natexlab{l}})}]{footneedmin} 
\bibinfo{note}{Of course, some energy minimization procedure would need to be applied to such packings in order to find the true local minima.}.

\bibitem[{\citenamefont{Royall and Williams}(2011)}]{royall11}
\bibinfo{author}{\bibfnamefont{C.~P.} \bibnamefont{Royall}} \bibnamefont{and}
  \bibinfo{author}{\bibfnamefont{S.~R.} \bibnamefont{Williams}},
  \bibinfo{journal}{J. Phys. Chem. B} \textbf{\bibinfo{volume}{115}},
  \bibinfo{pages}{7288} (\bibinfo{year}{2011}).
 
\bibitem[{foo({\natexlab{l}})}]{footM} 
\bibinfo{note}{Manoharan and collaborators have found a low fraction of Barlow packings (similar to that shown in Table \ref{tab:summarytab}) in colloids interacting via short-ranged depletion interactions \cite{perry12}.}  
 
 \bibitem[{\citenamefont{Mayer}(1942)}]{mayer42}
\bibinfo{author}{\bibfnamefont{J.~E.} \bibnamefont{Mayer}},
  \bibinfo{journal}{J. Chem. Phys.} \textbf{\bibinfo{volume}{10}},
  \bibinfo{pages}{629} (\bibinfo{year}{1942}). 
  
\bibitem[{\citenamefont{Perry, Meng, Dimiduk, Fung, and Manoharan}(2012)}]{perry12}
\bibinfo{author}{\bibfnamefont{R.}~\bibnamefont{Perry}}, \bibinfo{author}{\bibfnamefont{G.}~\bibnamefont{Meng}}, \bibinfo{author}{\bibfnamefont{T.} \bibnamefont{Dimiduk}},
\bibinfo{author}{\bibfnamefont{J.} \bibnamefont{Fung}}, \bibnamefont{and}
  \bibinfo{author}{\bibfnamefont{V.~N.} \bibnamefont{Manoharan}},
  \bibinfo{journal}{Faraday Discussions}, in press.
 
\bibitem[{foo({\natexlab{n}})}]{otherstudiesfoot} 
\bibinfo{note}{Other recent studies have examined formation of both stacking faults and 5-fold structures \cite{omalley03,zaccarrelli09,karayiannis11} in greater detail.}

\bibitem[{\citenamefont{Hoy and O'Hern}(2012)}]{hoy12}
\bibinfo{author}{\bibfnamefont{R.~S.} \bibnamefont{Hoy}} \bibnamefont{and}
  \bibinfo{author}{\bibfnamefont{C.~S.} \bibnamefont{O'Hern}},
  \bibinfo{journal}{Soft Matter} \textbf{\bibinfo{volume}{8}},
  \bibinfo{pages}{1215} (\bibinfo{year}{2012}).

\bibitem[{foo({\natexlab{o}})}]{footharrowell}
\bibinfo{note}{In constrast to hard sphere systems, where the critical nuclear
  size is $N \sim 100$ \cite{auer01,omalley03,harrowell10}.}

\bibitem[{\citenamefont{Harrowell}(2010)}]{harrowell10}
\bibinfo{author}{\bibfnamefont{P.}~\bibnamefont{Harrowell}},
  \bibinfo{journal}{J. Phys. Cond. Matt.} \textbf{\bibinfo{volume}{22}},
  \bibinfo{pages}{1} (\bibinfo{year}{2010}).

\bibitem[{\citenamefont{Cahn and Hilliard}(1958)}]{cahn58}
\bibinfo{author}{\bibfnamefont{J.~W.} \bibnamefont{Cahn}} \bibnamefont{and}
  \bibinfo{author}{\bibfnamefont{J.~E.} \bibnamefont{Hilliard}},
  \bibinfo{journal}{J. Chem. Phys.} \textbf{\bibinfo{volume}{28}},
  \bibinfo{pages}{258} (\bibinfo{year}{1958});
\bibinfo{author}{\bibfnamefont{J.~W.} \bibnamefont{Cahn}} \bibnamefont{and}
  \bibinfo{author}{\bibfnamefont{J.~E.} \bibnamefont{Hilliard}},
  \bibinfo{journal}{J. Chem. Phys.} \textbf{\bibinfo{volume}{31}},
  \bibinfo{pages}{688} (\bibinfo{year}{1959}).
  
 
   
\end{thebibliography}
\end{document}